\documentclass[fleqn,usenatbib]{mnras}

\usepackage{newtxtext,newtxmath}

\usepackage[T1]{fontenc}

\DeclareRobustCommand{\VAN}[3]{#2}
\let\VANthebibliography\thebibliography
\def\thebibliography{\DeclareRobustCommand{\VAN}[3]{##3}\VANthebibliography}

\usepackage{graphicx}	
\usepackage{amsmath}	

\usepackage{fix-cm}
\usepackage{xspace}
\usepackage{makecell}
\usepackage{pifont}
\usepackage{enumerate}  

\newcommand{\sfr}{\dot M_\star}
\newcommand{\ts}{\tau_\star}

\newcommand{\tdep}{\tau_\mathrm{d}}
\newcommand{\tdepi}{\tau_\mathrm{d}^{-1}}

\newcommand{\tsfh}{\tau_\mathrm{s}}
\newcommand{\tD}{t_\mathrm{D}}
\newcommand{\mdot}{\dot M}

\usepackage{xcolor}
\definecolor{pink}{rgb}{0.96, 0.76, 0.76}
\definecolor{aqua}{rgb}{0.22, 0.96, 0.93}
\definecolor{darkred}{rgb}{0.7, 0.0, 0.0}

\defcitealias{LeungNomoto2018}{LN18}
\defcitealias{Kobayashi2020}{K20}
\defcitealias{Gronow2021b}{G21}
\defcitealias{Seitenzahl2013}{S13} \defcitealias{ChieffiLimongi2004}{CL04}
\defcitealias{Kobayashi2006}{K06}
\defcitealias{Nugrid1}{Nugrid}
\defcitealias{Prantzos2018}{P18}

\title[Chemical evolution: metal-dependent yields]{Chemical separation of stellar populations: analytic solutions for chemical evolution models with metallicity-dependent yields}

\author[J.~L.~Sanders
]{Jason L. Sanders\thanks{jason.sanders@ucl.ac.uk (JLS)}
\\Department of Physics and Astronomy, University College London, London WC1E 6BT, UK
}

\date{Accepted XXX. Received YYY; in original form ZZZ}

\pubyear{2025}

\begin{document}
\label{firstpage}
\pagerange{\pageref{firstpage}--\pageref{lastpage}}
\maketitle

\begin{abstract}
Stellar abundances of elements with production channels that are metallicity-dependent (most notably aluminium) have provided an empirical route for separating different Galactic components. We present `single-zone' analytic solutions for the chemical evolution of galaxies when the stellar yields are metallicity-dependent. Our solutions assume a constant star formation efficiency, a constant mass-loading factor and that the yields are linearly dependent on the interstellar medium abundance (with the option of a saturation of the yields at high metallicity). We demonstrate how the metallicity dependence of the yields can be mathematically considered as a system-dependent delay time (approximately equal to the system's depletion time) that, when combined with system-independent delay times arising from stellar evolutionary channels, produces the separation of different systems based on their star formation efficiency and mass-loading factor. The utility of the models is highlighted through comparisons with data from the APOGEE spectroscopic survey.
We provide a comprehensive discussion of the chemical evolution models in the [Al/Fe]--[Mg/Fe] plane, a diagnostic plane for the separation of in-situ and accreted Galactic components. 
Extensions of the models are presented, allowing for the modelling of more complex behaviours largely through the linear combination of the presented simpler solutions. 
\end{abstract}



\section{Introduction}

\begin{figure*}
    \centering
    \includegraphics[width=\textwidth]{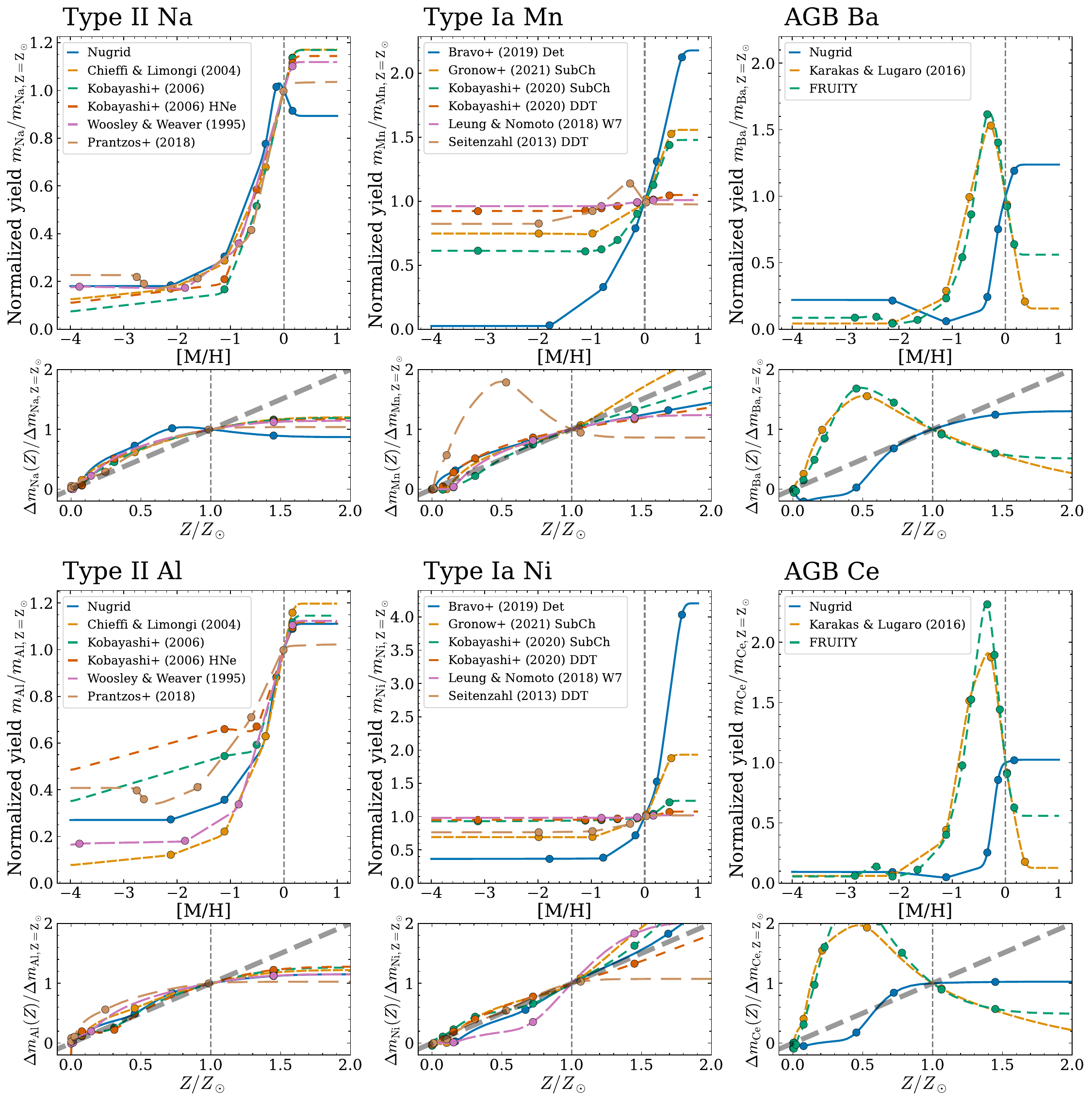}
    \caption{The net yield from different Type II (left column), Type Ia (centre) and AGB (right) models normalised by the production at solar metallicity. Each panel shows a different element. The Type II models are from \protect\cite{Nugrid1, Nugrid2, ChieffiLimongi2004, Kobayashi2006, WoosleyWeaver1995, Prantzos2018}, the Type Ia models from \protect\cite{Bravo2019, Gronow2021a, Gronow2021b, Kobayashi2020, LeungNomoto2018, Seitenzahl2013} and the AGB models from \protect\cite{Nugrid1, Nugrid2, KarakasLugaro, Lugaro2012, Karakas2018, Fruity1, Fruity2}. The dots show the metallicities at which models are available and the lines linearly interpolate between these and extrapolate using a constant value outside the available model range (some smoothing has been applied). The thicker dashed grey line in the top left panel shows a linear function.}
    \label{fig:metal_production}
\end{figure*}

Multidimensional chemical abundance patterns are a powerful tool both for the study of the chemical enrichment channels across the history of the Universe and for the study of galaxy formation \citep[e.g.][]{Parikh2019,Strom2021}. Chemical evolution models are a necessary part of the interpretation enabling characterisation both of the stellar processes contributing chemical elements and of the evolutionary properties of individual galaxies. Typically, these models come in two flavours: (i) purely numerical evolution of a set of enrichment equations using tables of nucleosynthetic yields from stellar evolution computations \citep[e.g.][]{Andrews2017,OMEGA,KobayashiChemicalEvolution,VICE}, and (ii) analytic models that make a number of simplifying assumptions to analytically integrate the enrichment equations. Whilst the former approach allows for easy incorporation of as many features as required, the latter approach is powerful as the models can be rapidly evaluated so are appropriate for exploring large regions of parameter space \citep{Sanders2021,FraserSchoenrich2022}, and the form of the equations enables a deeper mathematical understanding of the relative importance of different processes and assumptions.

One limitation of currently available analytic chemical evolution solutions \citep{Spitoni2017, Weinberg, Palicio2023} is the restriction to constant yields, i.e. after an infinite time, different channels (e.g. Type II supernovae, Type Ia supernovae, neutron star mergers etc.) return a fixed mass of element $x$ per unit star formation to the interstellar medium irrespective of its composition. Whilst such an approximation appears appropriate for several chemical abundances from several channels (e.g. O production from Type II supernovae), nucleosynthetic calculations demonstrate that the production of some elements is dependent on the abundance of other elements. These are often referred to as `secondary elements' though this terminology can be misleading as individual elements may be produced by multiple channels involving both primary and secondary processes. To model the distribution of secondary elements, the enrichment equations must depend upon the composition of the interstellar medium at some previous time, thus making them more complex.

There are several classic examples of secondary element yields. We display the results of numerical yield calculations for a number of these elements in Figure~\ref{fig:metal_production}. Typically any secondary dependence arises through the need for `seed' nuclei or the need for free neutrons, which are more abundant in higher metallicity environments. For example, the `odd-Z' elements, Na and Al (left column of Figure~\ref{fig:metal_production}), are predominantly produced in Type II supernovae, and their production is metallicity-dependent as it requires the excess neutrons provided by $^{22}$Ne produced via helium burning of $^{14}$N during the CNO cycle \citep{KobayashiChemicalEvolution}. Similarly, the production of some iron-peak elements, e.g. Mn, Ni and Co (central column of Figure~\ref{fig:metal_production}), in Type Ia supernovae is metallicity-dependent due to a higher neutron excess in more metal-rich systems that alters the relative abundance of iron-group elements from incomplete silicon burning \citep[e.g.][]{Gronow2021b}. Additionally, as all scenarios for Type Ia supernovae arise through binary interactions, the efficiency of Type Ia production itself could be a metallicity-dependent phenomenon both in terms of the properties of the initial binary population \citep{Moe2019} and in terms of the subsequent evolution. The production of elements in asymptotic giant branch (AGB) stars can be highly metallicity-dependent. Nitrogen is produced through the CN cycle of the CNO cycle so depends on the initial carbon abundance \citep{KarakasLattanzio2014}. Finally, the s-process in AGB stars requires neutrons to be successively captured onto (iron) seed nuclei. Increasing the metallicity in low metallicity AGB stars leads to more seed nuclei and the s-process enrichment is enhanced (right column of Figure~\ref{fig:metal_production}). However, at high metallicity, there are too many seed nuclei per free neutron bottlenecking the s-process and preventing it from reaching higher mass nuclei \citep[as highlighted in the case of Ce by][]{Weinberg2022}. To constrain any of these processes from chemical evolution modelling over a wide range of environments, metallicity dependence is a necessary part of the models.

The cosmic build-up of nitrogen is a particularly interesting example of secondary production \citep{Henry2000,Chiappini2003}. The dominant contribution to nitrogen production comes from intermediate-mass stars ($4-7M_\odot$) where secondary nitrogen yields scale with initial CNO abundances through the CN cycle. However, nitrogen also has a significant primary contribution from e.g. failed supernovae or massive rotating stars \citep{Prantzos2018,KobayashiChemicalEvolution}. Chemical evolution models of [N/O] vs. [O/H] measurements for populations of galaxies \citep[e.g.][]{Vincenzo2016,Berg2020,Schaefer2020} and populations of stars in the Milky Way \citep[e.g.][]{Johnson2023} have constrained the contribution and dependencies of these two production channels as well as other properties of the systems. This picture has been challenged by JWST observations of $z>6$ galaxies with surprisingly high [N/O] \citep{Bunker2023, Topping2024, Ji2025}, akin to that of some Milky Way globular cluster stars \citep{Belokurov2023}, which likely require additional primary contributions from very massive stars to explain \citep[e.g.][]{Vink2023}.

The abundances of elements produced predominantly through secondary processes also present an interesting route to the separation of different populations in the Galaxy. Traditionally, dwarf galaxies, or their remnants, have been chemically identified by their lower `knee' in the [Fe/H] vs. [$\alpha$/Fe] diagram arising from more effective outflows or lower star formation efficiencies in lower metallicity environments \citep{Weinberg}. The number of chemical abundances now measured per star from large spectroscopic surveys has seen investigations into superior abundance spaces for population separation. For example, the plane of [Al/Fe] vs. [Mg/Mn] has seen use as a tool for the classification of accreted and in-situ populations \citep{Hawkins2015, Das2020, Horta2021, BelokurovKravtsov2022, Conroy2022, OrtigozaUrdaneta2023, Horta2023, Fernandes2023, Feuillet2023, Vasini2023, Buckley2024, Li2024, Ernandes2025}. As discussed above, the yields of both Mn and Al are metallicity-dependent.
Typically, higher-mass galaxies are driven to higher metallicities, leading to enhanced production of elements with metallicity-dependent yields per unit iron, in particular [Al/Fe]. Once Type Ia supernovae begin contributing significantly, typically both [Al/Fe] and [Mg/Mn] start to decrease due to the increased production of Mn and Fe respectively. As a result, stellar populations typically lie along sequences in the [Al/Fe] vs. [Mg/Mn] plane, governed primarily by the metallicity reached at the onset of significant Type Ia enrichment.
However, although this behaviour is borne out of both empirical studies and numerical chemical evolution calculations, there is perhaps a more limited theoretical understanding of the chemical evolution pathways of galaxies in these ``secondary element planes''.

Here we present several analytic chemical evolution solutions with the inclusion of yields with linear metallicity dependence. Although more generalizable, our solutions are most simply expressed and most easily understood when the yields depend on the abundance of a `prompt' element, i.e. one solely produced in Type II supernova explosions, e.g. oxygen. We present our main chemical evolution modelling framework with the inclusion of metallicity-dependent yields in Section~\ref{section:model}. In Section~\ref{sec::specific}, we present analytic solutions for the chemical evolution given choices of star formation history. We discuss the behaviour of the tracks in different abundance planes in Section~\ref{sec::abundance_planes} with particular focus on the [Al/Fe] vs. [Mg/Fe] plane and its use for separating accreted and in situ populations. We conclude by discussing possible extensions and generalizations of our models in Section~\ref{sec::generalizations} before presenting our conclusions in Section~\ref{sec::conclusions}. We provide appendices which generalizes some of the presented results: the inclusion of a warm reservoir is presented in Appendix~\ref{appendix::warm}, the impact of recycling is investigated in Appendix~\ref{appendix::x_eq_y} and solutions using alternate delay-time distributions are given in Appendix~\ref{appendix::alternative_functional_forms_for_the_delay_time_distribution}. Throughout the exposition, we compare our solutions to spectroscopic data from the seventeenth data release of the APOGEE survey \citep{apogee}. We select all stars with \texttt{ASPCAPFLAG} $=0$, \texttt{STARFLAG} $=0$ and \texttt{MEMBERFLAG} $=0$. When plotting the abundances, we restrict ourselves to stars with uncertainties in $\mathrm{[X/Fe]}$<0.2 dex and with no flags in the relevant abundances.

\section{General formalism for chemical evolution models with metallicity-dependent yields}\label{section:model}

\emph{Derivation overview}: In this section and the next, we derive analytic solutions for chemical evolution models with metallicity-dependent yields. Before beginning, we map out the structure of the derivations, in particular where certain assumptions are made. We begin in equations~\eqref{eqn::ts}--\eqref{eqn::gx} by assuming constant star formation efficiency, $\ts^{-1}$, and constant mass-loading factor, $\eta$, with arbitrary enrichment channels, before specialising to channels with an exponential delay-time distribution from equation~\eqref{eqn::dtd}. The general solution is by the Laplace transform in equations~\eqref{eqn::laplace_defn}--\eqref{eqn::laplace}. From equation~\eqref{eqn:yield} onward, we assume a linear dependence of the yield on a prompt element (i.e. one produced only in Type II supernovae e.g. oxygen) and derive a simple equation to derive solutions with metallicity-dependent yields from their metallicity-independent counterparts in equation~\eqref{eqn::z1x_defn}. The general behaviour of these solutions is discussed in Section~\ref{sec::generic_behaviour}. Section~\ref{sec::specific}, presents the solutions for specific choices of star formation rate, including the case of yield saturation at high metallicity. A reference table for the timescales used in the models is given in Table~\ref{tab:parameters} and the key equations are summarized in Table~\ref{tab:equation_navigator} The Appendices provide complementary derivations: Appendix~\ref{appendix::x_eq_y} discusses how the recycling of elements can be considered in the same formalism, Appendix~\ref{appendix::warm} includes a warm reservoir in the solutions and Appendix~\ref{appendix::alternative_functional_forms_for_the_delay_time_distribution} explores alternative delay-time distributions.

Our modelling framework follows that of \cite{Weinberg}. We work with single-zone models assuming a linear Kennicutt-Schmidt law between the star-formation rate, $\sfr$, and the current gas mass $M$ as
\begin{equation}
\ts=M/\sfr,
\label{eqn::ts}
\end{equation}
for some constant star formation efficiency $\ts^{-1}$, and an outflow rate proportional to the star formation rate via a constant mass-loading factor $\eta$ defined as
\begin{equation}
\eta=\dot M_\mathrm{out}/\sfr.
\label{eqn::eta}
\end{equation}
The inflowing gas required to maintain this configuration is assumed to be pristine and the gas is assumed to be well mixed at all times (although as discussed below and in Appendix~\ref{appendix::warm}, generalization to enriched infall and multiple gas phases can be incorporated). Under this set of assumptions, the basic governing equation for the mass of element $x$, $M_x$, is
\begin{equation}
    \mdot_x +\tdepi M_x =\mathcal{G}_x,
    \label{eqn::main}
\end{equation}
where the depletion time is $\tdep\equiv\ts/(1+\eta)$ and $\mathcal{G}_x$ are the `source terms' representing the mass of element $x$ produced by stellar evolution and returned to the interstellar medium (ISM). We will primarily consider the production of \emph{new} elements rather than the recycling of elements. Appendix~\ref{appendix::x_eq_y} considers models for recycling and demonstrates that (for prompt return) the recycling can be considered as an additional contribution to the mass-loading factor $\eta\leftarrow\eta-r$ where $r$ is the returned fraction of mass to the ISM \citep{Weinberg}. In the following, any quoted values of $\eta$ should be considered as $\eta-r$ as the displayed models assume instantaneous recycling, but this can be relaxed using the results from Appendix~\ref{appendix::x_eq_y}.

Equation~\eqref{eqn::main} has the solution
\begin{equation}
    M_x(t) = \int_0^t\mathrm{d}t'\,\mathrm{e}^{-(t-t')/\tdep}\mathcal{G}_x(t'),
    \label{eqn::generalsoln}
\end{equation}
as $M_x(t=0)=0$, with the solution for the mass fraction as
\begin{equation}
Z_x(t)=\frac{M_x(t)}{M(t)}=\frac{1}{\ts}\frac{M_x(t)}{\sfr(t)}.
\label{eqn::generalsoln_Z}
\end{equation}
Generically, we can consider $\mathcal{G}_x$ as being a combination of a source term $S_x(t)$ describing all stellar products that will be returned from a population formed at time $t$ convolved with a delay-time distribution $\mathcal{D}(\tau)$ describing how long it takes those stellar products to reach the cold star-forming ISM such that
\begin{equation}
    \mathcal{G}_x(t)=\int_0^{t}\mathrm{d}t'\,\mathcal{D}(t-t')S_x(t').
\label{eqn::gx}
\end{equation}
Delayed return to the ISM can be due to either stellar evolution (e.g. in the case of products from Type Ia supernovae) or due to stellar products first entering a warm ISM phase and subsequently cooling to the cold star-forming phase. Here we limit ourselves to a single delay process for each production channel but in Appendix~\ref{appendix::warm}, we consider the more general case where some fraction of products from a delayed stellar evolutionary channel are first returned to a warm ISM phase and then cool to the cold ISM phase.

\citet[][see also \protect\citealt{Spitoni2017}]{Weinberg} demonstrate how analytic solutions of equations~\eqref{eqn::main} and~\eqref{eqn::gx} are possible under two assumptions:
\begin{enumerate}[(i)]
\item the delay-time distribution $\mathcal{D}(t)$ is exponential with minimum time $\tD$ and timescale $\tau_\mathrm{p}$ (`p' for production)
\begin{equation}
\mathcal{D}(t)=\frac{1}{\tau_\mathrm{p}}\mathrm{e}^{-(t-\tD)/\tau_\mathrm{p}}\Theta(t-\tD),
\label{eqn::dtd}
\end{equation}
where $\Theta(x)$ is the Heaviside step-function equal to $1$ for $x\geq0$ and $0$ for $x<0$,
\item and the source term is $S_x(t')=m_x \sfr$ for constant net yield $m_x$, the mass of newly produced element $x$ per unit mass of star formation.
\end{enumerate}
In this work, we will keep the assumed form for the delay-time distribution but consider relaxing the assumption of constant yields.

With these assumptions, equation~\eqref{eqn::main} is linear such that we can consider the contribution from several channels $S^i_x(t)$ by simply summing the solutions $M_x(t)$. Furthermore, more general delay-time distributions from a single channel can be considered as sums of the basic exponential delay-time distributions \cite{Weinberg,Sanders2021}. For example, the delay-time distribution for Type Ia supernovae is typically a power law which can be approximated as a sum of exponentials over a finite range in time. Appendix~\ref{appendix::alternative_functional_forms_for_the_delay_time_distribution} also discusses other possibly better physically-motivated choices of delay-time distribution that also yield analytic solutions but involve more complicated special functions making their computation and inspection more involved. We proceed to consider the solution from a single channel with a single exponential delay-time distribution acknowledging that any realistic model will likely be composed of a sum of these solutions.

\begin{table}
\caption{A reference table describing the timescales in our chemical evolution models.}
    \centering
    \renewcommand{\arraystretch}{1.5}
    \begin{tabular}{l|l|l}
         Symbol&Description&Equation\\
         \hline
         $\tau_\star$&Inverse of the star formation efficiency&\eqref{eqn::ts}\\
         $\tdep$&Depletion timescale, $\tau_\star/(1+\eta)$& \eqref{eqn::eta}\\
         $\tau_\mathrm{p}$&\makecell[tl]{Exponential delay timescale for elements \\produced through a stellar evolution channel}& \eqref{eqn::dtd}\\
         $t_\mathrm{D}$&\makecell[tl]{Minimum delay time for elements returned\\through some stellar evolutionary channel}& \eqref{eqn::dtd}\\
         $\tau_x$&\makecell[tl]{Secondary `timescale' governing importance\\of metallicity-dependent yields, $\tau_\star/(g_x m_\mathrm{O})$}& \eqref{eqn::tau_x_defn}\\
         $\tsfh$&\makecell[tl]{Star formation history timescale\\Exponential: $\sfr\propto\mathrm{e}^{-t/\tsfh}$ \\Linear-exponential: $\sfr\propto t\mathrm{e}^{-t/\tsfh}$}&\makecell[tl]{\\\eqref{eqn::exp_sfr_defn}\\\eqref{eqn::lexp_sfr_defn}}\\
         $\tau_{i,j}$&Shorthand for $(\tau_i^{-1}-\tau_j^{-1})^{-1}$& \eqref{eqn::tau_ij_defn}\\
         $t_\mathrm{c}$&\makecell[tl]{Saturation time: the time at which the linear\\metallicity dependence of the yields saturates}& \eqref{eqn::sx_saturated}\\
         $\tau_\mathrm{r}$&\makecell[tl]{Recycling timescale on which elements from\\AGB stars are returned}&\eqref{eqn::recycling1}\\
         $\tau_\mathrm{w}$&ISM cooling timescale (warm to cold)& \eqref{eq::gx_warm}
    \end{tabular}
    \label{tab:parameters}
\end{table}

\subsection{General solution by Laplace transform}

The simplest method for considering the solutions of equation~\eqref{eqn::main} is via the Laplace transform \citep[e.g.][]{Vicenzo2017}.
The Laplace transform of $f(t)$ is defined as
\begin{equation}
    u(s) \equiv \mathcal{L}_s\{f(t)\} = \int\mathrm{d}t\,\mathrm{e}^{-st}f(t).
    \label{eqn::laplace_defn}
\end{equation}
It satisfies three useful properties:
(i) differentiation becomes multiplication,
\begin{equation}
\mathcal{L}_s\{\dot f(t)\} = s u(s) - f(0),
\end{equation}
(ii) convolutions become products,
\begin{equation}
\mathcal{L}_s\Big\{\int_0^t\mathrm{d}t' f_1(t') f_2(t-t')\Big\} = u_1(s) u_2(s),
\label{eqn:laplace_property2}
\end{equation}
and (iii) shifts are replaced by exponential pre-factors,
\begin{equation}
    \mathcal{L}_s\Big\{f(t-t_\mathrm{D})\Theta(t-t_\mathrm{D})\Big\} = \mathrm{e}^{-t_\mathrm{D}s}u(s).
    \label{eqn:laplace_property3}
\end{equation}
These three properties allow the transformation of the governing differential equation~\eqref{eqn::main} into a purely algebraic function in Laplace space. This greatly simplifies the derivation of the solutions. An important Laplace transform for our solutions is that of an exponential function:
\begin{equation}
\mathcal{L}_s\Big\{\mathrm{e}^{-t/\tau}\Big\}=\frac{1}{s+\tau^{-1}}.
\end{equation}
If we denote the Laplace transform of $M_x(t)$ as
\begin{equation}
U_x(s)\equiv\mathcal{L}_s\{M_x(t)\}
\end{equation}
and say $M_x(t=0)=0$, then the Laplace transform of our governing equation \eqref{eqn::main} combined with the delay-time distribution (equation~\eqref{eqn::dtd}) is
\begin{equation}
U_x(s) = \frac{\mathrm{e}^{-t_\mathrm{D}s}}{\tau_\mathrm{p}}\frac{1}{s+\tau_\mathrm{d}^{-1}}\frac{1}{s+\tau_\mathrm{p}^{-1}}\mathcal{L}_s\Big\{S_x(t)\Big\}.
\label{eqn::laplace}
\end{equation}
Here we see an advantage of the Laplace transform: the convolution due to delayed return and the convolution arising from the depletion time due to star formation and outflows become products in the Laplace transform space. Indeed, repeated convolutions with exponential functions of timescales $\tau_i$ will become a product in Laplace space by equation~\eqref{eqn:laplace_property2} which can be separated into a sum by the identity
\begin{equation}
\prod_i\frac{1}{s+\tau_i^{-1}}=\sum_i\frac{\prod_{j\neq i}\tau_{j,i}}{s+\tau_i^{-1}}.
\label{eqn::identity1}
\end{equation}
This makes it obvious that if the base case of instantaneous production can be analytically solved for a specific choice of $\Gamma(s)$ then any additional delays coming from delayed return or the addition of a warm phase (see Appendix~\ref{appendix::warm} (and as we will see linear metallicity dependence) will also have analytic solutions. In the case of repeating factors, we can take the limit after the inverse Laplace transform.

In addition to being a useful mathematical device, the Laplace transform offers physical insight into how the system responds to different exponential driving terms. The variable $s$ represents a decay rate and $U_x(s)$ quantifies how strongly the system responds to components in $\mathcal{G}_x(t)$ that decay at that rate. In this sense, $U_x(s)$ acts like a transfer function, describing the system's sensitivity to enrichment sources with different temporal profiles. An advantage of the Laplace transform is that it has enabled us to find a general solution for $U_x(s)$ that is valid for \emph{any} star formation history. The specific form of $M_x(t)$ and $Z_x(t)$ can then be found through the inverse Laplace transform.

\subsection{Linear yield dependence on a solely prompt element}\label{section::x_neq_y}

In this work, our primary focus is on metallicity-dependent yields, $m_x(Z)$. We now consider the case where the yield is linearly dependent on the abundance of an element $y$. We define the constant of proportionality $g_x$ such that
\begin{equation}
    m_x(Z) = m_{x,0}(1+g_x Z_y),
\label{eqn:yield}
\end{equation}
so
\begin{equation}
    g_x = \frac{1}{m_{x,0}}\frac{\mathrm{d}m_x}{\mathrm{d}Z_y},
\label{eqn:gxdefn}
\end{equation}
where $m_{x,0}\equiv m_x(Z=0)$. Note that with this parametrisation, a pure secondary yield, one that vanishes at zero metallicity, corresponds to $m_{x,0}=0$. Only the product $g_xm_{x,0}$ enters the equations, so in this case, we introduce $b=g_xm_{x,0}$ and write $m_x(Z)=bZ_y$. However, in the following, we will keep the general primary plus secondary form in equation~\eqref{eqn:yield}.

For the general linear dependence of the yield in equation~\eqref{eqn:yield}, the source term, $S_x$, is given by
\begin{equation}
S_x(t) = m_{x,0}(1+g_x Z_y(t))\sfr(t) = m_{x,0}\Big(\sfr(t) + \frac{g_x}{\tau_\mathrm{\star}} M_y(t)\Big).
\label{eqn::s_x_linear_yield}
\end{equation}
This source term has a corresponding Laplace transform
\begin{equation}
\mathcal{L}_s\Big\{S_x(t)\Big\} = m_{x,0}\Big(\Gamma(s) + \frac{g_x}{\tau_\mathrm{\star}} U_y(s)\Big),
\label{eqn::laplace_source_Uy}
\end{equation}
where we have introduced 
\begin{equation}
    \Gamma(s)\equiv\mathcal{L}_s\Big\{\sfr(t)\Big\}.
\end{equation}
In Table~\ref{table:sfr_laplace} we give some physically-motivated example star formation laws and their corresponding Laplace transforms $\Gamma(s)$. These will form the basis of our detailed solutions in the next section.

\begin{table}
\caption{Example star formation laws $\sfr(t)$ with their corresponding Laplace transforms, $\Gamma(s)$. In the final row, $\tau_\mathrm{h}^{-1}\equiv\tau_1^{-1}+\tau_2^{-1}$.}
\centering
\begin{tabular}{c|c}
$\dot M_\star(t)$&$\Gamma(s)$\\
\hline
$1$&$s^{-1}$\\
$\mathrm{e}^{-t/\tau}$&$(s+\tau^{-1})^{-1}$\\
$t\mathrm{e}^{-t/\tau}$&$(s+\tau^{-1})^{-2}$\\
$(1-\mathrm{e}^{-t/\tau_1})\mathrm{e}^{-t/\tau_2}$&$(s+\tau_2^{-1})^{-1}-(s+\tau_\mathrm{h}^{-1})^{-1}$\\
\end{tabular}
\label{table:sfr_laplace}
\end{table}

The yield typically depends on the overall metallicity and detailed composition of the stars (as discussed in the Introduction). The overall metallicity evolution in a galaxy is a reflection of the operation of all production channels and even in the simple analytic framework outlined here, its functional dependence is complicated and we quickly lose the simplicity of the analytic approach. Here, we approximate the metallicity of the ISM as being purely the abundance of oxygen at any time i.e. $y=\mathrm{O}$. The approximation is justified as the typical dynamic range of $N_\mathrm{O}/N_\mathrm{Fe}$ is $2-3$ whilst the dynamic range of $N_\mathrm{O}/N_\mathrm{H}$ or $N_\mathrm{Fe}/N_\mathrm{H}$ can be $100-1000$. With this approximation, we therefore expect to capture the broad behaviour of any metallicity-dependent yields irrespective of the exact dependence. The reason we wish to make such an approximation is that the evolution of oxygen is very simple. Oxygen is a `prompt element' by which we mean it is almost wholly produced in Type II supernovae and Type II products can be assumed to be returned to the ISM nearly instantaneously (particularly compared to any other channel). For prompt elements, such as oxygen, we have that $\tau_\mathrm{p}\rightarrow0$ and $\tD=0$. This implies that $\mathcal{L}_s\Big\{S_\mathrm{O}(t)\Big\}=m_\mathrm{O}\Gamma(s)$ and the solutions to equation~\eqref{eqn::laplace} are given by:
\begin{equation}
U_\mathrm{O}(s) = m_\mathrm{O}\frac{\Gamma(s)}{s+\tau_\mathrm{d}^{-1}}.
\label{eqn::laplace_o}
\end{equation}
In this approximation, we can use equation~\eqref{eqn::laplace_source_Uy} with $y=\mathrm{O}$ to write the Laplace transform of the production of element $x$ from one channel as
\begin{equation}
\mathcal{L}_s\Big\{S_x(t)\Big\} = m_{x,0}\Big(\Gamma(s) + \frac{g_x}{\tau_\mathrm{\star}} U_\mathrm{O}(s)\Big).
\label{eqn::laplace_source_UO}
\end{equation}
In this case, the evolution for element $x$ is given by
\begin{equation}
U_x(s) = m_{x,0}\Gamma(s)\frac{\mathrm{e}^{-t_\mathrm{D}s}}{\tau_\mathrm{p}}\frac{1}{s+\tau_\mathrm{d}^{-1}}\frac{1}{s+\tau_\mathrm{p}^{-1}}\Big(1+\frac{\tau_x^{-1}}{s+\tau_\mathrm{d}^{-1}}\Big),
\label{eqn::x_neq_y}
\end{equation}
where we have introduced the \emph{secondary} `timescale'
\begin{equation}
    \tau_x = \frac{\tau_\mathrm{\star}}{g_x m_\mathrm{O}}.
\label{eqn::tau_x_defn}
\end{equation}
Although $\tau_x$ has units of time, it does not correspond directly to an enrichment timescale. It is instead the characteristic parameter that controls the importance of the metallicity-dependent term. As we will see, the actual timescale on which secondary production becomes important is controlled by $\tau_\mathrm{d}$ (and timescales in the star formation history). The pre-factor in equation~\eqref{eqn::x_neq_y} is the solution in the metallicity-independent yields case \citep[$g_x=0$, $\tau_x\rightarrow\infty$][]{Weinberg}. The additional term that arises when considering metallicity-dependent yields (i.e. the pre-factor times the second term in the final bracket) can be simply expressed as the derivative of this metallicity-independent case. If we denote
\begin{equation}
U^0_x(s)=\lim_{g_x\rightarrow0}U_x(s),
\end{equation}
then
\begin{equation}
U_x(s) = \Big(1-\frac{1}{\tau_x}\frac{\partial}{\partial\tau_\mathrm{d}^{-1}}\Big)U^0_x(s).
\end{equation}
Now to solve for the evolution of the mass of element $x$, we take the inverse Laplace transform
\begin{equation}
M_x(t)=\mathcal{L}^{-1}_t\Big\{U_x(s)\Big\}
\end{equation}
which, due to the linearity of the Laplace transform, implies
\begin{equation}
M_x(t)=\Big(1-\frac{1}{\tau_x}\frac{\partial}{\partial\tau_\mathrm{d}^{-1}}\Big)M_x^0(t)
\end{equation}
where $M_x^0(t)$ is the solution for the mass of element $x$ in the metallicity-independent yields case. Now finally, we find the evolution of the mass \emph{fraction} of element $x$ by using equation~\eqref{eqn::generalsoln_Z}, $Z_x = M_x/(\tau_\star \sfr(t))$, and noting that $\sfr(t)$ is independent of $\tau_\mathrm{d}$ so
\begin{equation}
Z_x(t) = \Big(1-\frac{1}{\tau_x}\frac{\partial}{\partial\tau_\mathrm{d}^{-1}}\Big)Z^0_x(t),
\label{eqn::simple_metal}
\end{equation}
where
\begin{equation}
Z^0_x(t)=\lim_{g_x\rightarrow0}Z_x(t).
\label{eqn::z0x_defn}
\end{equation}
We define the additional piece arising from metallicity dependence of the yields as $Z^1_x(t)$ where
\begin{equation}
Z^1_x(t) \equiv -\frac{1}{\tau_x}\frac{\partial Z^0_x}{\partial\tau_\mathrm{d}^{-1}},
\label{eqn::z1x_defn}
\end{equation}
such that the full solution for the mass fraction is the sum
\begin{equation}
Z_x(t)\equiv Z_x^0(t)+Z_x^1(t).
\label{eqn::simple_metal_sum}
\end{equation}
Here we have taken the independent parameters to be $\tau_\mathrm{d}$ and $\tau_\star$, so the derivative $\partial/\partial\tau_\mathrm{d}^{-1}$ is always taken at constant $\tau_\star$. Equivalently, as $\tau_\mathrm{d}^{-1}=(1+\eta)/\tau_\star$, this means we are taking derivatives with respect to $\eta$ whilst holding $\tau_\star$ constant.

Equation~\eqref{eqn::z1x_defn} is the key result from this section: we have a convenient expression to find via differentiation the {metallicity-dependent} solutions directly from the metallicity-independent solutions derived by \cite{Weinberg}. We will see how this relation gives rise to useful analytic expressions for specific star formation history choices. In computational implementations, the derivative could be evaluated via auto-differentiation. $Z^0_x(t)$ describes the results of primary processes whilst $Z^1_x(t)$ describes the results of secondary processes. A general process with both primary and secondary contributions is obtained by the linear sum of $Z^0_x$ and $Z^1_x$, as in equation~\eqref{eqn::simple_metal_sum}. In Appendix~\ref{appendix::warm} we show that the inclusion of a warm ISM reservoir that adds an additional cooling timescale to the equations produces a similar simple expression for $Z^1_x(t)$ in terms of derivatives of both the depletion time \emph{and} the cooling time.

The presented formulation can also be used for modelling the recycling of elements, i.e. no production, if we choose $x=y$ in equation~\eqref{eqn::s_x_linear_yield}. We work through this case explicitly in Appendix~\ref{appendix::x_eq_y} and, as already discussed, show that in the prompt limit ($\tau_\mathrm{p}\rightarrow0$ and $t_\mathrm{D}=0$), it is equivalent to an adjustment of the mass-loading factor, $\eta\leftarrow\eta-r$ where $r$ is the recycled mass fraction \citep{Weinberg}. More generally, with the general source term in equation~\eqref{eqn::s_x_linear_yield} we can consider modelling the \emph{gross} yields from different stellar channels where $g_x$ includes the contribution from metallicity-dependent production \emph{and} recycled elements. However, in what follows we do not explicitly consider recycling but instead use an effective mass-loading factor which implicitly includes the effects of recycling.

\subsection{Generic behaviour of the solutions}\label{sec::generic_behaviour}
Before going on to discuss specific solutions to the presented equations, we discuss the general behaviour of the solutions when metallicity-dependent yields are included.
The depletion timescale, $\tdep$, controls how rapidly elements are removed from the ISM through star formation and outflows. If $\tdep$ increases, elements are removed more slowly leading to an increased build-up arising from the return of stellar products. This implies that the derivative in equation~\eqref{eqn::z1x_defn}, $\partial Z^0_x(t)/{\partial \tau_\mathrm{d}^{-1}}=-\tau_\mathrm{d}^2\partial Z^0_x(t)/{\partial \tau_\mathrm{d}}$, is negative which in turn implies that the additional term in equation~\eqref{eqn::simple_metal} will lead to an enhancement in the mass fraction for $g_x>0$. This intuitively makes sense -- if the yield increases with metallicity and the metallicity increases over time, the mass fraction will increase faster than in the metal-independent case. We can see this more mathematically by differentiating equation~\eqref{eqn::generalsoln} to find
\begin{equation}
    Z_x(t)=Z_x^0(t)+\frac{1}{\tau_x\tau_\star\sfr(t)}\int_0^t\mathrm{d}t'(t-t')\mathrm{e}^{-(t-t')/\tau_\mathrm{d}}\mathcal{G}_x(t'),
\end{equation}
where the second term is clearly always positive as $t'\leq t$.

A useful limit to consider is the prompt limit ($t_\mathrm{D}=0$, $\tau_\mathrm{p}=0$) with metallicity-dependent yields. This is applicable to Al production from Type II supernovae. In this case, the metallicity-dependent part of equation~\eqref{eqn::x_neq_y} (pre-factor times the second term in the bracket) reduces to the same form as the metallicity-independent yield solution (equation~\eqref{eqn::x_neq_y} with $\tau_x^{-1}=0$) with the replacement $\tau_\mathrm{p}=\tau_\mathrm{d}$. Therefore, in this limit, the solutions from \cite{Weinberg} apply directly with the replacement of the delay time with the depletion time. This demonstrates explicitly that a delayed return is mathematically identical to secondary production (under our modelling assumptions) as discussed by \cite{Henry2000} and \cite{Chiappini2003}. Note, however, that there is a system-to-system difference in the secondary production case as the effective delay time is the depletion time, $\tau_\mathrm{d}=\tau_\star/(1+\eta)$. Via equation~\eqref{eqn::laplace_o}, the depletion time can be considered as the timescale to enrich the system to a required metallicity. This system-dependent delay time depends on the star formation efficiency, $\tau_\star^{-1}$, and mass-loading factor, $\eta$. However, in the case of an element from a channel with a delay time resulting from stellar evolution (e.g. $s$ process production from AGB stars), the delay time, $\tau_\mathrm{p}$, is independent of the system. We therefore see the power of secondary elements for separation of populations from different systems: they behave as elements with a delayed production channel that crucially depends on the system properties.

\begin{table}
    \caption{Net yields for use with the models \protect\citep[in units of the solar abundance from][]{Asplund2009}. We adopt the \protect\cite{Kroupa2001} initial mass function for computation of the Type II yields and the number of Type Ia supernovae per stellar mass formed as $2.2\times10^{-3}M_\odot^{-1}$ from \protect\cite{MaozMannucci2012}. The Type II yields are from \protect\citet[][CL04]{ChieffiLimongi2004}, \protect\citet[][K06, w/H includes hypernovae]{Kobayashi2006}, \protect\citet[][Nugrid]{Nugrid1,Nugrid2} and \protect\citet{Prantzos2018}. The Type Ia yields are from \protect\citet[][LN18, their W7 model]{LeungNomoto2018}, \protect\citet[][K20 are DDT models and K20sCh are sub-Chandrasekhar models]{Kobayashi2020}, \protect\citet[][S13]{Seitenzahl2013} and \protect\citet[][G21]{Gronow2021b}.
    $m_{x,0}$ is computed at [M/H] $=-1.8$ for elements with metallicity-dependent yields (Na, Al, Mn, Ni) and [M/H] $=-0.1$ otherwise. $g_x$ are computed via equation~\eqref{eqn:finite_diff}.
    The key quantity for metallicity dependence of the yields is $g_xm_\mathrm{O}$, obtained by multiplying the gradient rows with the Type II oxygen yield (top row).}
    \centering
    \renewcommand{\arraystretch}{1.2}
    \begin{tabular}{l|ccccc}
\textbf{Type II}& \citetalias{ChieffiLimongi2004} & \citetalias{Kobayashi2006} & \citetalias{Kobayashi2006}w/H & \citetalias{Nugrid1} & \citetalias{Prantzos2018}
\\\hline
$m^\mathrm{II}_\mathrm{O,0}/Z_\mathrm{O,\odot}$ &$2.49$&$2.86$&$1.59$&$0.95$&$1.52$\\
$m^\mathrm{II}_\mathrm{Mg,0}/Z_\mathrm{Mg,\odot}$ &$1.57$&$1.81$&$1.14$&$0.50$&$0.50$\\
$m^\mathrm{II}_\mathrm{Na,0}/Z_\mathrm{Na,\odot}$ &$0.73$&$0.59$&$0.41$&$0.19$&$0.21$\\
$m^\mathrm{II}_\mathrm{Al,0}/Z_\mathrm{Al,\odot}$ &$0.40$&$1.32$&$0.91$&$0.09$&$0.24$\\
$m^\mathrm{II}_\mathrm{Fe,0}/Z_\mathrm{Fe,\odot}$ &$0.95$&$0.64$&$0.54$&$1.20$&$0.61$\\
\\
$g^\mathrm{II}_\mathrm{Na}Z_\mathrm{O,\odot}$ &$4.44$&$7.64$&$5.70$&$5.18$&$4.69$\\
$g^\mathrm{II}_\mathrm{Al}Z_\mathrm{O,\odot}$ &$6.41$&$1.07$&$0.65$&$2.81$&$1.90$\\
\noalign{\vskip 15pt}
\textbf{Type Ia}& \citetalias{LeungNomoto2018}W7 & \citetalias{Kobayashi2020} & \citetalias{Seitenzahl2013} & \citetalias{Gronow2021b}sCh & \citetalias{Kobayashi2020}sCh \\
\hline
$m^\mathrm{Ia}_\mathrm{Mn,0}/Z_\mathrm{Mn,\odot}$ &$2.42$&$2.28$&$1.63$&$0.70$&$0.06$\\
$m^\mathrm{Ia}_\mathrm{Fe,0}/Z_\mathrm{Fe,\odot}$ &$1.37$&$1.46$&$1.26$&$0.99$&$1.11$\\
$m^\mathrm{Ia}_\mathrm{Ni,0}/Z_\mathrm{Ni,\odot}$ &$1.95$&$2.07$&$1.72$&$0.67$&$0.28$\\
\\
$g^\mathrm{Ia}_\mathrm{Mn}Z_\mathrm{O,\odot}$ &$0.05$&$0.10$&$0.34$&$0.37$&$0.67$\\
$g^\mathrm{Ia}_\mathrm{Ni}Z_\mathrm{O,\odot}$ &$0.01$&$0.05$&$0.33$&$0.47$&$0.06$
    \end{tabular}
    \label{tab:yields}
\end{table}

\begin{figure*}
    \centering
    \includegraphics[width=\textwidth]{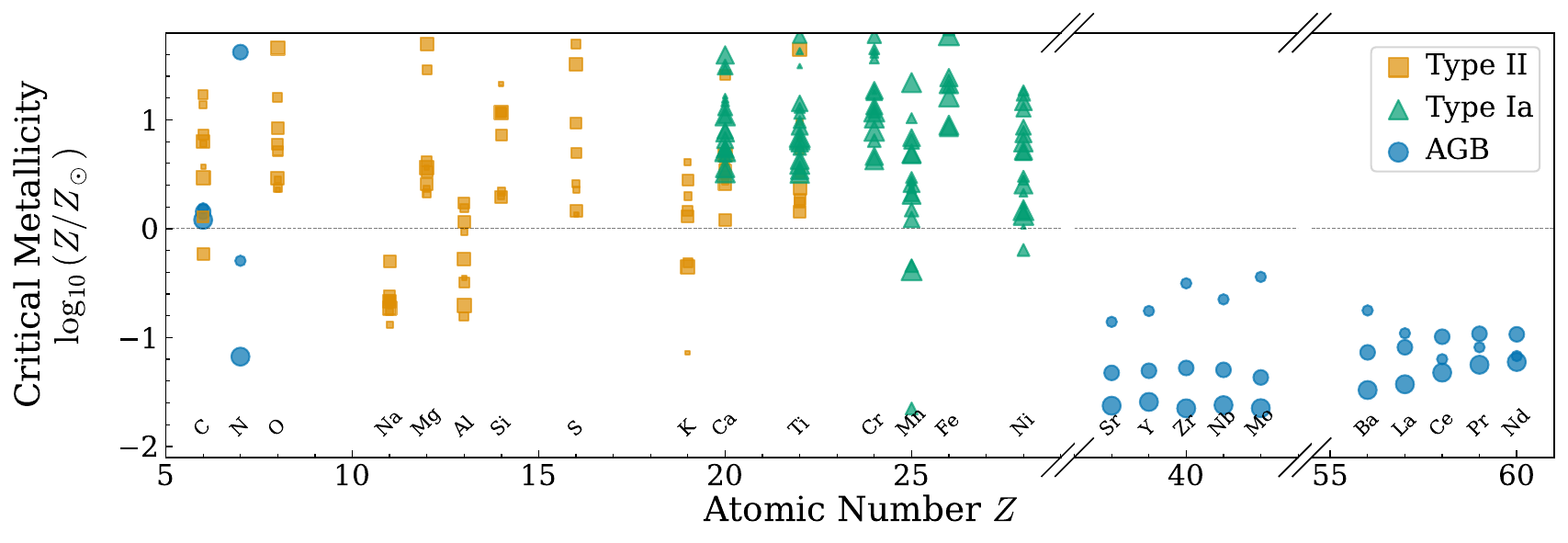}
    \caption{Critical metallicity above which metallicity dependence of yields is `significant' (computed as the inverse of the yield metallicity gradient from finite differencing models at $[\mathrm{M/H}]=-1.8$ and $[\mathrm{M/H}]=-0.1$). At the critical metallicity, the metal dependence of the yields for element X approximately gives rise to shifts of $\log_{10}[\mathrm{X}/\mathrm{O}]=0.43$ (see equation~\eqref{eqn:delta_metal_shift}). The Type II models (orange squares) are
    Nugrid \protect\citep{Nugrid1, Nugrid2}, \protect\citet[][without and with hypernovae]{Kobayashi2006}, \protect\cite{ChieffiLimongi2004}, \protect\citet[][no rotation, $150\,\mathrm{km/s}$ and $300\,\mathrm{km/s}$]{LimongiChieffi2018}, \protect\cite{Prantzos2018} and \protect\cite{WoosleyWeaver1995}, the Type Ia models (green triangles) are
    \protect\citet[][$M=(0.8,0.9,1,1.1)M_\odot$]{Gronow2021b}, \protect\citet{Kobayashi2006}, \protect\cite{Seitenzahl2013}, \protect\cite{Bravo2019}, \protect\citet[][$M=(0.9,1,1.1,1.2)M_\odot$]{Kobayashi2020}, \protect\cite[][$M=(1.3,1.33,1.37)M_\odot$]{Kobayashi2020}, \protect\citet[][default, low density, high density and W7]{LeungNomoto2018}, \protect\cite{LeungNomoto2020} and \protect\cite{Shen2018}, and the AGB models (blue circles) are Nugrid \protect\citep{Nugrid1, Nugrid2}, \protect\citeauthor{KarakasLugaro} \protect\citep{Lugaro2012, KarakasLugaro, Karakas2018} and FRUITY \protect\citep{Fruity1, Fruity2}, all in order of increasing point size. Note in particular that the critical metallicity for Al falls exactly at the separation between the typical metallicity of in-situ Milky Way stars (i.e. solar) and those in dwarf galaxies $\mathrm{[M/H]}\approx-1$.}
    \label{fig:crit_metallicity}
\end{figure*}

For weak metallicity dependence of the yields when $\tau_\mathrm{d}\ll|\tau_x|$ (and also significantly shorter than any timescales in the star formation history, $\sfr$), we use the approximation $(t-t')\mathrm{exp}(-(t-t')/\tau_\mathrm{d}) \approx \tau_\mathrm{d}\mathrm{exp}(-(t-t')/\tau_\mathrm{d})$ to find
\begin{equation}
Z_x(t) \approx Z_x^0(t)(1+\tau_\mathrm{d}/\tau_x).
\label{eqn::weak_limit_asymptote}
\end{equation}
The enrichment is a scaled version of the metal-independent solution. In this limit, the mass fraction of a metallicity-independent prompt species (e.g. oxygen) will rapidly approach an equilibrium value $Z_x^0(t)\approx m_{x,0}\tau_\mathrm{d}/\tau_\star$. We can therefore approximate the small change in the abundance ratio of a prompt element $x$ to another metallicity-independent prompt element (e.g. oxygen) due to the metal independence of the yields as
\begin{equation}
\Delta [\mathrm{X}/\mathrm{O}]\approx\frac{1}{\ln 10} g_x Z_\mathrm{O}.
\label{eqn:delta_metal_shift}
\end{equation}
As expected, the solutions will only show significant metallicity dependence if the yields do. This discussion demonstrates that the critical threshold for metallicity dependence to be important is $\tau_x=\tau_\mathrm{d}$ or in terms of metallicity $Z_\mathrm{O,\mathrm{crit}}\approx1/g_x$. By equation~\eqref{eqn:delta_metal_shift}, this is the critical metallicity at which the change in abundance due to the metallicity yields is $1/\ln10\approx0.43$.

In Fig.~\ref{fig:crit_metallicity}, we have plotted this critical metallicity $[\mathrm{M/H}]_\mathrm{crit}=-\log_{10}g_x Z_{\mathrm{O},\odot}$ for a range of elements from a range of net yield calculations for Type II, Type Ia and AGB channels. The gradients $g_x Z_{\mathrm{O},\odot}$ have been approximated as
\begin{equation}
\begin{split}
g_x Z_{\mathrm{O},\odot}&=\frac{\mathrm{d}(m_x/m_{x,0})}{\mathrm{d}(Z/Z_\odot)}\\&\approx\frac{m_x([\mathrm{M/H}]=-0.1)/m_x([\mathrm{M/H}]=-1.8)-1}{10^{-0.1}-10^{-1.8}}.
\end{split}
\label{eqn:finite_diff}
\end{equation}
We see in general significant scatter across different yield calculations (in some part due to fundamentally different setups e.g. Type Ia detonations vs. deflagrations), but broad trends are apparent. For Type II models, the critical metallicity for the majority of elements is super-solar except for the odd-Z elements, Na, Al and K. In particular, the critical metallicity for Al falls somewhere $-1\lesssim[\mathrm{M}/\mathrm{H}]_\mathrm{crit}\lesssim0$, hence why Al is useful for separating accreted dwarfs with $[\mathrm{M}/\mathrm{H}]\approx-1$ from the in-situ Milky Way stars. We also see that the most metallicity-sensitive Type Ia yields are Mn and Ni, also known for their value in separating near-Chandrasekhar from sub-Chandrasekhar models. Finally, the AGB yields show that the metallicity dependence of all $s$-process element yields is important even for low metallicity. We give numerical values for some $m_{x,0}$ and $g_x$ for different elements from Type II and Type Ia supernovae from several studies in Table~\ref{tab:yields}. Note $m_{x,0}$ and $g_x$ have been calculated from net yields i.e. the new products from the stellar channels.

We have motivated the broad dependence on timescales from inspection of the general forms of the solution. These considerations required taking limits of the equations and ignoring complexity introduced by delayed return of the elements and star formation history. To inspect the dependence of the formalism on these properties, we turn to inspecting analytic solutions for specific choices of the star formation history.\\

\section{Specific choices of star formation history}\label{sec::specific}

\begin{table}
    \caption{A reference table for the chemical evolution model results for different choices of star formation history in different limits. Note the constant star formation law can be recovered from the exponential by the replacements $\tau_{\mathrm{d},\mathrm{s}}\rightarrow\tau_\mathrm{d}$, $\tau_{\mathrm{p},\mathrm{s}}\rightarrow\tau_\mathrm{p}$ and $\mathrm{e}^{\tD/\tau_\mathrm{s}}\rightarrow1$.}
    \centering
    \begin{tabular}{l|ccc}
         \textbf{Cases}&Exponential&Linear-exponential\\
         \hline
         General&\eqref{eqn::general_z}, \eqref{eqn:const_sfr_curlyZ}&\eqref{eqn:leq_general_z},\eqref{eqn:leq_general_curlyz}\\
         Saturated&\eqref{eqn:trunc_exp}&\eqref{eqn:trunc_lexp}\\
         Early-time&\eqref{eqn::exp_early}&\eqref{eqn::lexp_early}\\
         Late-time&\eqref{eqn::exp_late}&\eqref{eqn::exp_late}\\
         \hline
         $\tau_\mathrm{p}=0$&\eqref{eqn::exp_zerotp}&\eqref{eqn::lexp_tp0}\\
         $\tau_\mathrm{p}=0$ early-time&\eqref{eqn::exp_zerotp_early}&\eqref{eqn::lexp_tp0_early}\\
    \end{tabular}
    \label{tab:equation_navigator}
\end{table}

\begin{figure*}
    \centering
    \includegraphics[width=\textwidth]{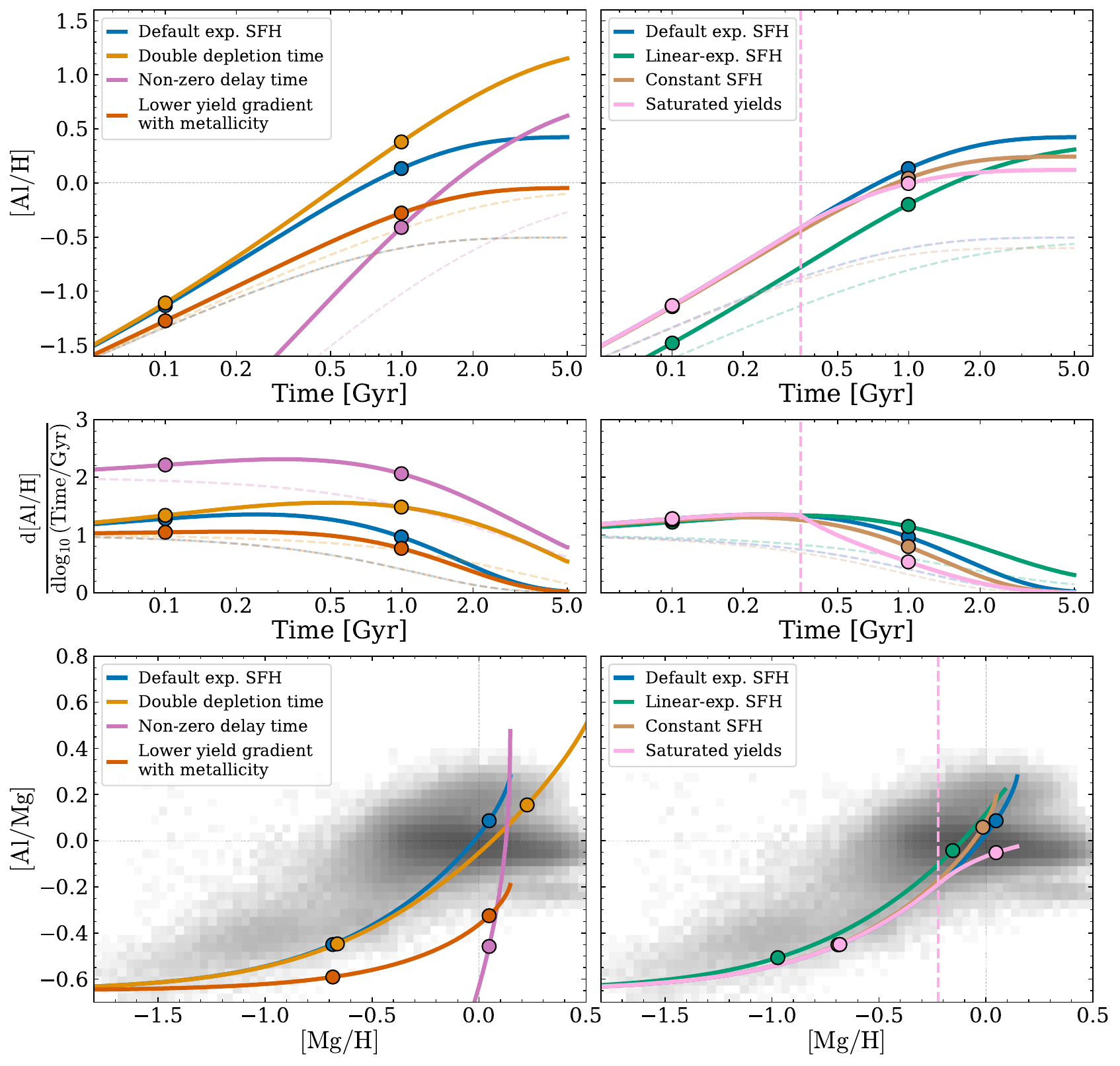}
    \caption{Evolution of abundances for Mg, a representative prompt element with metallicity-independent yields and Al, a representative prompt element with metallicity-dependent yields. The \textbf{top row} shows evolution with time for [Al/H] in solid lines (and in faint dashed when metallicity dependence is switched off, $g_\mathrm{Al}=0$).
    The \textbf{middle row} shows the derivative of [Al/H] with respect to the logarithm of time. The \textbf{bottom row} shows the evolution of the abundance ratio [Al/Mg] with [Mg/H]. The dots correspond to times of $0.1$ and $1\,\mathrm{Gyr}$ and the model is generated for a total of $5\,\mathrm{Gyr}$. Each line shows a different model configuration: the default setup (blue) is an exponential star formation history with $\tau_\star=1\,\mathrm{Gyr}$, $\eta=1$ and $\tau_\mathrm{s}=2.5\,\mathrm{Gyr}$ (appropriate for the Milky Way thick disc) and is shown in both left and right panels; \textbf{left}: orange doubles the depletion time $\tau_\mathrm{d}=\tau_\star/(1+\eta)$ by altering $\eta$); dark pink introduces a non-zero delay time for Al of $\tau_\mathrm{p}=1.5\,\mathrm{Gyr}$; red lowers the metallicity gradient of the yields, $g$, by a factor $4$; \textbf{right}: green uses a linear-exponential star formation history; brown uses a constant star formation history and pink saturates the growth of the yields with metallicity at $t_\mathrm{c}=0.35\,\mathrm{Gyr}$. The greyscale shows the density of the APOGEE DR17 dataset.}
    \label{fig:example_models}
\end{figure*}
With the general framework of chemical evolution models with metallicity-dependent yields developed, we now present analytic solutions in the case where the yields depend solely on a prompt element e.g. oxygen for specific choices of the star formation rate: exponential, constant (a limiting case of the exponential) and linear-exponential. We will also inspect the solutions in the early and late time regimes. The key derived equations are summarised in Table~\ref{tab:equation_navigator}.

\subsection{Exponential star formation rate}\label{sub:exponential_star_formation_rate}
We begin with the exponential star formation rate,
\begin{equation}
    \sfr\propto \mathrm{e}^{-t/\tau_\mathrm{s}},
    \label{eqn::exp_sfr_defn}
\end{equation}
as other results can be obtained from this case either through taking the limit of $\tau_\mathrm{s}\rightarrow\infty$ for the constant star formation rate or through differentiation with respect to $\tau_\mathrm{s}^{-1}$ in the case of a linear exponential star formation rate, $t\mathrm{e}^{-t/\tau_\mathrm{s}}$. Furthermore, this simple star formation rate can be used as a base function to compute more general star formation rates through linear combinations \citep[e.g.][]{Weinberg2023}.

As shown in Table~\ref{table:sfr_laplace}, the Laplace transform for this choice of star formation rate is $\Gamma(s)=(s+\tau_\mathrm{s}^{-1})^{-1}$ meaning equation~\eqref{eqn::x_neq_y} is purely a product of factors like $(s+\tau^{-1})^{-1}$. As shown in equation~\eqref{eqn::identity1}, these products can be separated out as a sum over the factors $(s+\tau^{-1})^{-1}$, each of which has an exponential as the inverse Laplace transform (Table~\ref{table:sfr_laplace}). Therefore, for the exponential star formation rate, the solution in the metal-independent case is
\begin{equation}
Z_x^0(t) =
Z_{x,\mathrm{eq}}^0\Big(1-\mathrm{e}^{-\Delta t/\tau_{\mathrm{d},\mathrm{s}}}
-\frac{\tau_{\mathrm{d},\mathrm{p}}}{\tau_{\mathrm{d},\mathrm{s}}}\Big[
\mathrm{e}^{-\Delta t/\tau_{\mathrm{p},\mathrm{s}}}-\mathrm{e}^{-\Delta t/\tau_{\mathrm{d},\mathrm{s}}}\Big]\Big),
\label{eqn::general_z}
\end{equation}
for $\Delta t=t-t_\mathrm{D}>0$ otherwise $Z_x^0$ is zero. Here we adopt the notation similar to that of \cite{Weinberg} of
\begin{equation}
\tau_{i,j}^{-1}\equiv\tau_i^{-1}-\tau_j^{-1}.
\label{eqn::tau_ij_defn}
\end{equation}
We have introduced the equilibrium metallicity, $Z_{x,\mathrm{eq}}^0$,
\begin{equation}
Z_{x,\mathrm{eq}}^0= m_{x,0}\mathrm{e}^{\tD/\tau_\mathrm{s}}\frac{\tau_{\mathrm{d},\mathrm{s}}\tau_{\mathrm{p},\mathrm{s}}}{\tau_\mathrm{p}\tau_\star},
\label{eqn::equilibrium}
\end{equation}
which is the asymptotic limit of $Z_x^0$ as $t\rightarrow\infty$ provided $\tau_\mathrm{p}<\tau_\mathrm{s}$ and $\tau_\mathrm{d}<\tau_\mathrm{s}$. The latter of these constraints imposes a positive inflowing gas mass. Using equation~\eqref{eqn::z1x_defn}, this implies that the additional part in the metallicity-dependent case is
\begin{equation}
Z^1_x(t) = \frac{\tau_{\mathrm{d},\mathrm{s}}}{\tau_x}\Big(Z_x^0-
Z_{x,\mathrm{eq}}^0\frac{\tau_{\mathrm{d},\mathrm{p}}^2}{\tau_{\mathrm{d},\mathrm{s}}\tau_{\mathrm{p},\mathrm{s}}}
\Big[\mathrm{e}^{-\Delta t/\tau_{\mathrm{p},\mathrm{s}}}-\Big(1+\frac{\Delta t}{\tau_{\mathrm{d},\mathrm{p}}}\Big)\mathrm{e}^{-\Delta t/\tau_{\mathrm{d},\mathrm{s}}}\Big]\Big),
\label{eqn:const_sfr_curlyZ}
\end{equation}
again with the proviso that $\Delta t>0$ otherwise $Z^1_x=0$.
At early times ($\Delta t\rightarrow0$), we find
\begin{equation}
\begin{split}
Z_x^0(t)&=\tfrac{1}{2}\frac{m_{x,0}\mathrm{e}^{t_\mathrm{D}/\tau_\mathrm{s}}}{\tau_\star\tau_\mathrm{p}}\Delta t^2 + \mathcal{O}(\Delta t^3);\\
Z^1_x(t)&=\tfrac{1}{6}\frac{m_{x,0}\mathrm{e}^{t_\mathrm{D}/\tau_\mathrm{s}}}{\tau_\star\tau_\mathrm{p}\tau_x}\Delta t^3 + \mathcal{O}(\Delta t^4).
\end{split}
\label{eqn::exp_early}
\end{equation}
This implies the ratio of the enrichment of an element produced by purely secondary processes to that of an element produced by purely primary processes in the case where both come from the same delayed channel grows linearly in time at early times.
In the limit $t\rightarrow\infty$ provided $\tau_\mathrm{p}<\tau_\mathrm{s}$ and $\tau_\mathrm{d}<\tau_\mathrm{s}$, we find
\begin{equation}
\lim_{t\rightarrow\infty}
Z_x(t)\rightarrow Z_{x,\mathrm{eq}}^0(1+\tau_{\mathrm{d},\mathrm{s}}/\tau_x),
\label{eqn::exp_late}
\end{equation}
i.e. the additional term in equation~\eqref{eqn:const_sfr_curlyZ} tends to zero, the result is independent of the additional delay time, $\tau_\mathrm{p}$ and the metallicity dependence leads to an enhancement $(1+\tau_{\mathrm{d},\mathrm{s}}/\tau_x)$ of the equilibrium mass fraction (provided $g_x>0$, compare to equation~\eqref{eqn::weak_limit_asymptote}).

In the prompt return limit of $\tD=0$ and $\tau_\mathrm{p}\rightarrow0$, we have $Z_{x,\mathrm{eq}}^0=m_{x,0}\tau_{\mathrm{d},\mathrm{s}}/\tau_\star$ and the solutions become
\begin{equation}
\begin{split}
Z^0_x(t) &= \frac{m_{x,0} \tau_{\mathrm{d},\mathrm{s}}}{\tau_\star}(1-\mathrm{e}^{-t/\tau_{\mathrm{d},\mathrm{s}}});\\
Z^1_x(t) &= m_{x,0} \frac{\tau_{\mathrm{d},\mathrm{s}}^2}{\tau_\star\tau_x}(1-\mathrm{e}^{-t/\tau_{\mathrm{d},\mathrm{s}}}-(t/\tau_{\mathrm{d},\mathrm{s}})\mathrm{e}^{-t/\tau_{\mathrm{d},\mathrm{s}}}).
\end{split}
\label{eqn::exp_zerotp}
\end{equation}
In this case, the early-time behaviour is
\begin{equation}
\begin{split}
Z^0_x(t)&=\frac{m_{x,0}}{\tau_\star}t + \mathcal{O}(t^2);\\
Z^1_x(t)&=\tfrac{1}{2}\frac{m_{x,0}}{\tau_\star\tau_x}t^2 + \mathcal{O}(t^3),
\end{split}
\label{eqn::exp_zerotp_early}
\end{equation} and again the ratio of two elements produced by the same channel with one produced by purely secondary processes and one produced by purely primary processes grows linearly in time at early times.

In Fig.~\ref{fig:example_models}, we plot examples of these models using a default setup with $t_\star=1\,\mathrm{Gyr}$, $\eta=1$, $\tau_\mathrm{s}=2.5\,\mathrm{Gyr}$ (suitable for modelling the Milky Way thick disc),
$\log_{10}(m_\mathrm{Mg,0}^\mathrm{II}/Z_\mathrm{Mg,\odot})=+0.35$,
$\log_{10}(m_\mathrm{Al,0}^\mathrm{II}/Z_\mathrm{Al,\odot})=-0.3$ (Table~\ref{tab:yields}) and $g_\mathrm{Al}m_\mathrm{O}=12$. We see the linear rise at early times, steepening once the metallicity dependence of the yields becomes significant and plateauing towards the equilibrium values at late times. The [Al/Mg] vs. [Mg/H] ratio is flat at low [Mg/H] when the metallicity dependence of the yields is unimportant before rising in an approximately quadratic fashion. Note that whilst the models keep rising at high [Mg/H], the APOGEE data appear to slightly flatten. We will discuss abundance ratios further later. Doubling the depletion time means more elements are retained and the system is driven to higher metallicities earlier. Introducing a delay time to Al produces lower [Al/H] at earlier times and a steepening of the evolution, whilst decreasing the yield gradient $g_\mathrm{Al}$ naturally decreases the rise of [Al/H] with time.

\subsubsection{Comparison with other literature models}

The solutions derived here bear a similarity to previously considered models in the literature.
\cite{Henry2000} computed analytic models of elements from prompt secondary (and tertiary) processes under the assumption that $(1-\alpha)\sfr$ of the star formation is locked up in stellar remnants and the inflow rate $\dot M_\mathrm{inflow}$ is linearly proportional to the star formation rate: $\alpha a=\dot M_\mathrm{inflow}/\sfr$. This is a different assumption to our starting premise of a linear Kennicutt-Schmidt law that, whilst quite common due to its simplicity, is perhaps less physically motivated as the star formation rate is likely unrelated directly to the inflow rate \citep{Maiolino2019}. However, the two assumptions are compatible if the star formation rate is exponential such that $\dot M=(\tau_\star/\tau_\mathrm{s})\sfr$ and $\dot M=-\alpha(1-a)\sfr$. This means our equations in the exponential case coincide with \cite{Henry2000} if $a=\tau_\mathrm{d}/\tau_{\mathrm{d,s}}$ and $\alpha=\tau_\star/\tau_\mathrm{d}$.

Another interesting limiting case to consider is the classic closed box model with no outflows or inflows and only prompt return of stellar products. In this case, under our linear Kennicutt-Schmidt law assumption, the star formation rate is required to be exponential with $\tau_\star=\tau_\mathrm{s}=\tau_\mathrm{d}$. The known solution of the mass fraction of an element from purely primary processes, such as oxygen, is $Z_\mathrm{O}(t)=m_\mathrm{O} t/\tau_\mathrm{s}$ and the solution for the mass fraction of an element with secondary production linearly dependent on the abundance of oxygen is
\begin{equation}
Z_x(t) = \frac{m_{x,0}}{\tau_\mathrm{s}}\Big(t+\frac{1}{2}\frac{g_xm_\mathrm{O}}{\tau_\mathrm{s}}t^2\Big)=\frac{m_{x,0}}{m_\mathrm{O}}Z_\mathrm{O}\Big(1+\frac{1}{2}g_xZ_\mathrm{O}\Big),
\end{equation}
as given by \cite{Tinsley1980}.

\subsection{Constant star formation rate}
A second case to consider is that of constant star formation rate, which under our assumed Kennicutt-Schmidt law implies constant gas mass. The solution in this case can be found from the exponential star formation rate in the limit $\tau_\mathrm{s}\rightarrow\infty$. This means we use equations~\eqref{eqn::general_z} and~\eqref{eqn:const_sfr_curlyZ} with $\tau_{\mathrm{d},\mathrm{s}}\rightarrow\tau_\mathrm{d}$ and $\tau_{\mathrm{p},\mathrm{s}}\rightarrow\tau_\mathrm{p}$ and $\mathrm{e}^{\tD/\tau_\mathrm{s}}\rightarrow1$. In Fig.~\ref{fig:example_models}, we show the constant star formation rate case alongside the exponential case with $\tau_\mathrm{s}=2.5\,\mathrm{Gyr}$. Increasing $\tau_\mathrm{s}$ means higher gas densities are required to sustain star formation at late times so the equilibrium metallicity is lowered.

For the prompt limit of the constant star formation rate case ($\tD=0$ and $\tau_\mathrm{p}\rightarrow0$), we recover the `extreme infall case' discussed by \cite{Tinsley1980} who considered the limit of no outflows ($\eta=0$) and constant gas mass. The equations from \cite{Tinsley1980} are recovered by making the identifications that the yields, $y_x$, defined by \cite{Tinsley1980} as the mass of new metals ejected per unit mass locked in stars are $y_x\equiv m_{x,0} \tau_\mathrm{d}/\tau_\star$ and $M_\star/M=t/\tau_\mathrm{d}$.

\subsection{Linear-exponential star formation rate}
The exponential star formation rate cases are mathematically simple but somewhat unphysical as they assume the star formation instantly switches on at some time. We therefore consider the slightly more physical case of a linear-exponential star formation history, $t\mathrm{e}^{-t/\tau_\mathrm{s}}$. We will first demonstrate that this case can be derived from the exponential star formation history case via differentiation of the solutions.

For a general star formation law, $\sfr^a$, parametrized by $\theta$ with corresponding solution for mass of element $x$ equal to $M_x^a$, we can differentiate equation~\eqref{eqn::main} with respect to $\theta$ to show that if the star formation rate is $\sfr^b=\partial\sfr^a/\partial\theta$, the solution for the mass of element $x$ will be $M_x^b=\partial M_x^a/\partial \theta$. This means the mass fraction for element $x$ will be
\begin{equation}
    Z_x^b=\frac{M_x^b}{\sfr^b}=\frac{\partial M_x^a}{\partial\theta}\Big(\tau_\star\frac{\partial\sfr^a}{\partial\theta}\Big)^{-1}=\frac{\partial (Z_x^a \sfr^a)}{\partial\theta}\Big(\frac{\partial\sfr^a}{\partial\theta}\Big)^{-1}.
\label{eqn::derivs_relation_sfr}
\end{equation}
Therefore, in the case where
\begin{equation}
    \sfr\propto t\mathrm{e}^{-t/\tau_\mathrm{s}},
    \label{eqn::lexp_sfr_defn}
\end{equation}
we can derive the results from the previous exponential case by first recognising that $\sfr\propto t\mathrm{e}^{-t/\tau_\mathrm{s}} = -\partial(\mathrm{e}^{-t/\tau_\mathrm{s}})/\partial \tau_\mathrm{s}^{-1}$and then, using equation~\eqref{eqn::derivs_relation_sfr} to find
\begin{equation}
Z_x = -\frac{1}{t\mathrm{e}^{-t/\tau_\mathrm{s}}}\frac{\mathrm{d}}{\mathrm{d}\tau_\mathrm{s}^{-1}}\Big(Z_{x,\mathrm{exp}} \mathrm{e}^{-t/\tau_\mathrm{s}}\Big),
\label{eqn::exp_to_linearexp}
\end{equation}
where $Z_{x,\mathrm{exp}}$ is the solution for the exponential star formation rate (equations~\eqref{eqn::general_z} and~\eqref{eqn:const_sfr_curlyZ}). Therefore, we find that for the metallicity-independent case, we have
\begin{equation}
Z_x^0(t) = Z_{x,\mathrm{eq}}^0\frac{\tau_{\mathrm{p},\mathrm{s}}}{t}\Big(\frac{\Delta t}{\tau_{\mathrm{p},\mathrm{s}}}+\frac{\tau_{\mathrm{d},\mathrm{p}}}{\tau_{\mathrm{d},\mathrm{s}}}\mathrm{e}^{-\Delta t/\tau_{\mathrm{p},\mathrm{s}}}
-\frac{\tau_{\mathrm{d},\mathrm{s}}\tau_{\mathrm{d},\mathrm{p}}}{\tau_{\mathrm{p},\mathrm{s}}^2}\mathrm{e}^{-\Delta t/\tau_{\mathrm{d},\mathrm{s}}}
-\Big[1+\frac{\tau_{\mathrm{d},\mathrm{s}}}{\tau_{\mathrm{p},\mathrm{s}}}\Big]
\Big),
\label{eqn:leq_general_z}
\end{equation}
and
\begin{equation}
\begin{split}
Z^1_x(t) = \frac{\tau_{\mathrm{d},\mathrm{s}}}{\tau_x}\Big(Z_x^0
&+Z_{x,\mathrm{eq}}^0\frac{\tau_{\mathrm{d},\mathrm{p}}}{t}
\Big[
\frac{\tau_{\mathrm{d},\mathrm{p}}}{\tau_{\mathrm{d},\mathrm{s}}}\mathrm{e}^{-\Delta t/\tau_{\mathrm{p},\mathrm{s}}}
\\&-\frac{\tau_{\mathrm{d},\mathrm{p}}}{\tau_{\mathrm{p},\mathrm{s}}}
\mathrm{e}^{-\Delta t/\tau_{\mathrm{d},\mathrm{s}}}
\Big\{
1+\frac{\tau_{\mathrm{d},\mathrm{s}}}{\tau_{\mathrm{d},\mathrm{p}}}+\frac{\Delta t}{\tau_{\mathrm{d},\mathrm{p}}}
\Big\}
-\frac{\tau_{\mathrm{d},\mathrm{s}}}{\tau_{\mathrm{d},\mathrm{p}}}
\Big]
\Big),
\end{split}
\label{eqn:leq_general_curlyz}
\end{equation}
again with the proviso that $\Delta t>0$ otherwise $Z^0_x=Z^1_x=0$. The constant $Z_{x,\mathrm{eq}}^0$ is the same as for the exponential case (given in equation~\eqref{eqn::equilibrium}) which is the value $Z_x^0(t)$ tends to at late times provided $\tau_\mathrm{p}<\tau_\mathrm{s}$ and $\tau_\mathrm{d}<\tau_\mathrm{s}$ such that the late time limit given in equation~\eqref{eqn::exp_late} is also valid in the linear exponential star formation rate case.

For early times $\Delta t\ll t_\mathrm{D} \ll \tau_\mathrm{p,s}, \tau_\mathrm{d,s}$, we find that
\begin{equation}
\begin{split}
Z_x^0(t)&=\tfrac{1}{6}\frac{m_{x,0}}{\tau_\star\tau_\mathrm{p}t_\mathrm{D}}\Delta t^3 + \mathcal{O}(\Delta t^4);\\
Z^1_x(t)&=\tfrac{1}{24}\frac{m_{x,0}}{\tau_\star\tau_\mathrm{p}t_\mathrm{D}\tau_x}\Delta t^4 + \mathcal{O}(\Delta t^5).
\end{split}
\label{eqn::lexp_early}
\end{equation}
For times longer than the minimum delay time, $t_\mathrm{D}$, but still shorter than the other timescales in the problem (i.e. $t_\mathrm{D}\ll \Delta t \ll \tau_\mathrm{p,s}, \tau_\mathrm{d,s}$, we find
\begin{equation}
\begin{split}
Z_x^0(t)&=\tfrac{1}{6}\frac{m_{x,0}}{\tau_\star\tau_\mathrm{p}}\Delta t^2 + \mathcal{O}(\Delta t^3);\\§
Z^1_x(t)&=\tfrac{1}{24}\frac{m_{x,0}}{\tau_\star\tau_\mathrm{p}\tau_x}\Delta t^3 + \mathcal{O}(\Delta t^4),
\end{split}
\label{eqn::lexp_early_td0}
\end{equation}
which are also appropriate solutions for $t_\mathrm{D}=0$. In the prompt return limit of $t_\mathrm{D}=0$ and $\tau_\mathrm{p}\rightarrow0$, we have
\begin{equation}
\begin{split}
Z_x^0(t) &= \frac{m_{x,0}\tau_{\mathrm{d},\mathrm{s}}}{\tau_\star}\Big(1-\tau_{\mathrm{d},\mathrm{s}}\frac{1-\mathrm{e}^{-t/\tau_{\mathrm{d},\mathrm{s}}}}{t}\Big);\\
Z^1_x(t) &= \frac{m_{x,0}\tau_{\mathrm{d},\mathrm{s}}^2}{\tau_\star\tau_x}\Big(1+\mathrm{e}^{-t/\tau_{\mathrm{d},\mathrm{s}}}-2\tau_{\mathrm{d},\mathrm{s}}\frac{1-\mathrm{e}^{-t/\tau_{\mathrm{d},\mathrm{s}}}}{t}\Big).
\end{split}
\label{eqn::lexp_tp0}
\end{equation}
For early times $t\ll|\tau_\mathrm{d,s}|$, these look like
\begin{equation}
\begin{split}
Z_x^0(t)&=\tfrac{1}{2}\frac{m_{x,0}}{\tau_\star}t + \mathcal{O}(t^2);\\
Z^1_x(t)&=\tfrac{1}{6}\frac{m_{x,0}}{\tau_\star\tau_x}t^2 + \mathcal{O}(t^3).
\end{split}
\label{eqn::lexp_tp0_early}
\end{equation}

Fig.~\ref{fig:example_models} compares the linear-exponential models to the other choices of star formation history. We see the linear-exponential star formation history has a slower rise to the equilibrium metallicity but in the [Al/Mg] vs. [Mg/H] plane the models are very similar although note the models reach different ([Mg/H], [Al/Mg]) points at fixed times (shown by the dots).

\subsection{Saturating linear growth}\label{sub:truncating_linear_growth}
As shown in Fig.~\ref{fig::example_tracks} and already noted by \cite{Tinsley1980}, solutions where the yields from prompt channels increase linearly with the metallicity have the generic property of producing $[\mathrm{X}/\mathrm{Y}]$ vs. $[\mathrm{Y}/\mathrm{H}]$ that increases faster than linear, for element $\mathrm{X}$ with secondary process contributions and element $\mathrm{Y}$ with purely primary process contributions. The data, however, appear to show a slight flattening in e.g. [Al/Mg] and [Na/Mg] around solar metallicity (Fig.~\ref{fig:example_models}). For the Milky Way data, this could be a result of superposition of different populations (which could be captured using a multi-zone model) or a genuine feature of the yields. There is some suggestion of the latter in Fig.~\ref{fig:metal_production} but note that in this figure, the yields are extrapolated using a constant value for high metallicities. The saturation of the yields at high metallicity is also physically motivated: the secondary production of some elements is highly sensitive to the ratio of free neutrons to seed nuclei. This quantity is metallicity-dependent and at high metallicity, there are many seed nuclei which `soak up' the free neutrons bottlenecking the successive capture of neutrons required to build up to the higher mass elements.
This discussion motivates the need to go beyond the simple \emph{linear} dependence of the yields on oxygen abundance to account for this complexity.

We, therefore, consider a \emph{saturated} source term given by
\begin{equation}
\begin{split}
    S^\mathrm{c}_x(t)&\equiv m_{x,0}\Big(1+g_x\min (Z_\mathrm{O}(t),Z_\mathrm{c})\Big)\sfr(t)\\&=S_x(t)-m_{x,0}g_x(Z_\mathrm{O}(t)-Z_\mathrm{O}(t_\mathrm{c}))\Theta(t-t_\mathrm{c})\sfr(t),
    \end{split}
\label{eqn::sx_saturated}
\end{equation}
where the metallicity-dependent part saturates at $Z=Z_\mathrm{c}$ occurring at time $t=t_\mathrm{c}$. If one chooses the free parameter as $Z_\mathrm{c}$, the solution for $t_\mathrm{c}$ must, in general, be found numerically.

\subsubsection{Exponential star formation rate}
In the exponential star formation rate case, the saturation time is related analytically to the saturation metallicity as \begin{equation}
t_\mathrm{c}=\tau_{\mathrm{s},\mathrm{d}}\ln(1+Z_\mathrm{c}/m_{x,0}\tau_\star/\tau_{\mathrm{s},\mathrm{d}}).
\end{equation}
For this specific exponential star formation rate only, we can write part of the second term in equation~\eqref{eqn::sx_saturated} as
\begin{equation}
\Big(Z_\mathrm{O}(t)-Z_\mathrm{O}(t_\mathrm{c})\Big)\sfr(t) = Z_\mathrm{O}(t-t_\mathrm{c})\sfr(t-t_\mathrm{c})\mathrm{e}^{-t_\mathrm{c}/\tau_{\mathrm{d}}}.
\end{equation}
This can be checked by substituting in the solution for $Z_\mathrm{O}(t)$ from equation~\eqref{eqn::exp_zerotp}. We now use the property of the Laplace transform in equation~\eqref{eqn:laplace_property3} to write
\begin{equation}
\mathcal{L}_s\Big\{Z_\mathrm{O}(t-t_\mathrm{c})\sfr(t-t_\mathrm{c})\Theta(t-t_\mathrm{c})\Big\}=\tau_\star^{-1}\mathrm{e}^{-st_\mathrm{c}}U_\mathrm{O}(s),
\end{equation}
so the Laplace transform of equation~\eqref{eqn::sx_saturated} is
\begin{equation}
\mathcal{L}_s\Big\{S^\mathrm{c}_x(t)\Big\}=
\mathcal{L}_s\Big\{S_x(t)\Big\}-\mathrm{e}^{-t_\mathrm{c}/\tau_\mathrm{d}}\frac{m_{x,0}g_x}{\tau_\star}\mathrm{e}^{-st_\mathrm{c}}U_\mathrm{O}(s).
\end{equation}
When this is inserted into equation~\eqref{eqn::laplace}, the second term is of the form that arises from the metallicity-dependent yields in e.g. equation~\eqref{eqn::laplace_source_UO} but with a shift of $t_\mathrm{c}$ from the $\mathrm{e}^{-s t_\mathrm{c}}$ factor (see equation~\eqref{eqn:laplace_property3}). This means the mass of element $x$ is
\begin{equation}
    M_x^\mathrm{c}(t) = M_x(t)-\mathrm{e}^{-t_\mathrm{c}/\tau_\mathrm{d}}M_x^1(t-t_\mathrm{c})\Theta(t-t_\mathrm{c}),
\end{equation}
where $M_x(t)$ is the full solution without saturation and $M_x^1(t)=Z_x^1(t)M(t)$ is the metallicity-dependent part of the mass of element $x$.
Therefore, the solution for the abundance fraction in this specific exponential star formation history case is
\begin{equation}
    Z^\mathrm{c}_x(t)=\frac{M^\mathrm{c}_x(t)}{\tau_\star \sfr(t)}=Z_x(t)-\mathrm{e}^{-t_\mathrm{c}/\tau_{\mathrm{d},\mathrm{s}}}Z^1_x(t-t_\mathrm{c})\Theta(t-t_\mathrm{c}),
\label{eqn:trunc_exp}
\end{equation}
where $Z_x(t)$ is the solution without the saturation.
Again the constant star formation rate limit can be simply found by taking $\tau_\mathrm{s}\rightarrow\infty$.

We show the saturated yields model for the exponential star formation history in Fig.~\ref{fig:example_models}. $t_\mathrm{c}$ is set at $0.35\,\mathrm{Gyr}$ corresponding to [Mg/H]$\approx-0.2\,\mathrm{dex}$. After this, [Al/H] plateaus and [Al/Mg] vs. [Mg/H] turns over, better capturing the distribution of the APOGEE data.

\subsubsection{Linear-exponential star formation rate}
In the case of the linear-exponential star formation rate, the saturation time can be related to the saturation metallicity as
\begin{equation}
    t_\mathrm{c} = \tau_{\mathrm{d}, \mathrm{s}}(W(A\mathrm{e}^A)-A); \quad\quad A\equiv \Big(\frac{\tau_\star Z_\mathrm{c}}{m_{x,0}\tau_{\mathrm{d}, \mathrm{s}}}-1\Big)^{-1},
\end{equation}
where we have introduced the Lambert $W$ function defined as the solution of $W(z)\mathrm{e}^{W(z)}=z$. The result in this case for the evolution of some element $x$ is
\begin{equation}
    Z^\mathrm{c}_x(t)=Z_x(t)-\mathrm{e}^{-t_\mathrm{c}/\tau_{\mathrm{d},\mathrm{s}}}\Big(1-\frac{t_\mathrm{c}}{t}\Big)\Big[Z^1_x(t-t_\mathrm{c})
+\frac{\tau_{\mathrm{d},\mathrm{s}}^2f}{\tau_x t_\mathrm{c}}Z_x^0(t-t_\mathrm{c})
    \Big]\Theta(t-t_\mathrm{c}),
\label{eqn:trunc_lexp}
\end{equation}
where
\begin{equation}
f = \mathrm{e}^{t_\mathrm{c}/\tau_{\mathrm{d},\mathrm{s}}}-\frac{t_\mathrm{c}}{\tau_{\mathrm{d},\mathrm{s}}}-1.
\end{equation}
This can be derived from equation~\eqref{eqn:trunc_exp} using the derivative relation in equation~\eqref{eqn::derivs_relation_sfr}.

\subsection{Summary}
We have given specific analytic results for chemical evolution models with linear metallicity-dependent yields for three related star formation histories: exponential, constant and linear-exponential. The results presented in this section are summarised in Table~\ref{tab:equation_navigator}. As shown in equation~\eqref{eqn::simple_metal_sum}, a general solution for the mass fraction of element $x$ is formed by summing $Z_x^0(t)$, the part arising from the metallicity-independent part of the yields (equation~\eqref{eqn::z0x_defn}), and $Z_x^1(t)$, the part arising from the metallicity-dependent part of the yields (equation~\eqref{eqn::z1x_defn}). Table~\ref{tab:equation_navigator} gives the equation numbers for these two terms for the exponential and linear-exponential star formation rates in the general case, the case of saturated yields, the prompt limit ($\tau_\mathrm{p}=0$) and the early and late-time limits. In Section~\ref{sec::users} we provide a fuller ``User's Guide'' for implementing the models presented in this paper.

\section{Models of abundance planes}\label{sec::abundance_planes}
We have seen how the basic analytic solutions behave as a function of time. Richer behaviour is observed when one begins comparing different elements to each other, particularly if there is more than one production channel for the elements.

\subsection{Abundance ratios from individual channels}\label{sub:abundance_ratios_from_individual_channels}
We now consider the ratio of the mass fractions of elements $x$ and $y$, assuming each is produced by a single nucleosynthetic channel. We take $x$ to be produced entirely through secondary processes and $y$ to come solely from a prompt channel with purely primary production e.g. $y$ could be oxygen or magnesium.

\begin{figure*}
\includegraphics[width=\textwidth]{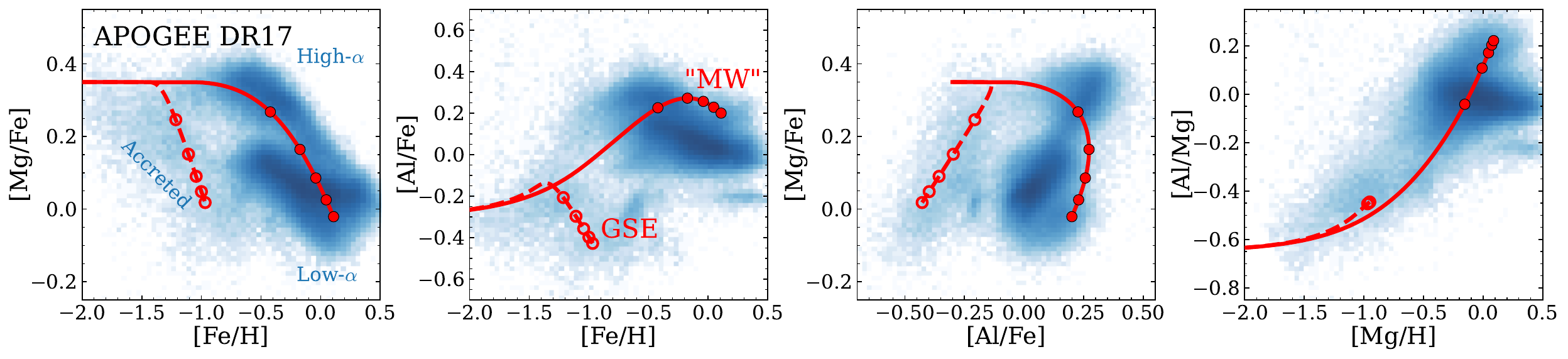}
\caption{Example solutions for our analytic chemical evolution models plotted over APOGEE DR17 data (blue logarithmic density). The solid red line is an example ``Milky Way'' track and the dashed red line is an example Gaia-Sausage Enceladus (GSE) track. The dots (circles) are spaced by $1\,\mathrm{Gyr}$.}
\label{fig::example_tracks}
\end{figure*}
First, we consider the case where $x$ is also prompt e.g. Al from Type II supernovae. The ratio at early times ($t\ll |\tau_\mathrm{d,s}|$) goes like
\begin{equation}
\lim_{t\rightarrow0}\frac{Z_x}{Z_y}=\frac{m_{x,0}}{m_{y,0}}\Big(1+C\frac{t}{\tau_x}\Big)=\frac{m_{x,0}}{m_{y,0}}\Big(1+C\frac{\tau_\star}{\tau_x}\frac{Z_y}{m_{y,0}}\Big),
\label{eqn::low_t_ratio}
\end{equation}
where $C=1/2$ for the exponential case and $1/3$ for the linear-exponential case,
and tends to the equilibrium value of
\begin{equation}
\lim_{t\rightarrow\infty}\frac{Z_x}{Z_y}=\frac{m_{x,0}}{m_{y,0}}\Big(1+\frac{\tau_\mathrm{d,s}}{\tau_x}\Big)=\frac{m_{x,0}}{m_{y,0}}\Big(1+\frac{\tau_\star}{\tau_x}\frac{Z_y}{m_{y,0}}\Big),
\label{eqn::high_t_ratio}
\end{equation}
provided $\tau_\mathrm{d,s}>0$. The extrapolation of the early time solution underpredicts the late time limit demonstrating that the ratio $[\mathrm{X}/\mathrm{Y}]$ with $[\mathrm{Y}/\mathrm{H}]$ will steepen at late times \citep{Tinsley1980}. As discussed above and as already noted by \cite{Tinsley1980}, these solutions possibly mismatch the data (see Fig.~\ref{fig:example_models}) as typically the ratio of elements with a secondary component to their yields, e.g. Al, to those produced mostly by primary processes, e.g. Mg, flattens at high metallicity. This observed behaviour could reflect the fundamental behaviour of stellar yields with metallicity and motivated our generalization to a saturated dependence in Section~\ref{sub:truncating_linear_growth}.

We note that we can approximately stitch together the two solutions in equations~\eqref{eqn::low_t_ratio} and~\eqref{eqn::high_t_ratio} to find the abundance ratio at some arbitrary time as
\begin{equation}
    \frac{Z_x}{Z_y}\approx \frac{m_{x,0}}{m_{y,0}}\Big(1+\frac{Ct/\tau_x}{1+Ct/\tau_\mathrm{d,s}}\Big).
\label{eqn::stitch}
\end{equation}
We also observe from equations~\eqref{eqn::exp_zerotp} and~\eqref{eqn::lexp_tp0} that the equations for the metallicity-independent part of the exponential and linear-exponential star formation history models in the prompt limit are of the form
\begin{equation}
Z_x^0(t)\equiv \frac{m_{x,0}\tau_{\mathrm{d,s}}}{\tau_\star}f(t/\tau_{\mathrm{d,s}}),
\end{equation}
for some function $f(x)$ from which the derivative relation in equation~\eqref{eqn::z1x_defn} implies
\begin{equation}
Z^1_x(t) = \frac{m_{x,0}\tau_{\mathrm{d,s}}^2}{\tau_\star \tau_x}\Big[
f(t/\tau_{\mathrm{d,s}})-(t/\tau_{\mathrm{d,s}})f'(t/\tau_{\mathrm{d,s}})\Big],
\end{equation}
where the square bracket is importantly just a function of $t/\tau_{\mathrm{d,s}}$.
We can then write
\begin{equation}
    \frac{Z_x}{Z_y}=\frac{m_{x,0}}{m_{y,0}}\Big[1+g_xm_\mathrm{O}\frac{\tau_{\mathrm{d,s}}}{\tau_\star}h\Big(\frac{\tau_\star}{\tau_{\mathrm{d,s}}}\frac{Z_y(t)}{m_{y,0}}\Big)\Big],
\label{eqn::z_x_z_y_with_z_y}
\end{equation}
for some function $h(x)$. This means in these models $Z_x/Z_y$ is not solely a function of $Z_y$ for prompt elements $x$ and $y$, but different system parameter combinations $\tau_\star/\tau_{\mathrm{d,s}}=1+\eta-\tau_\star/\tsfh$ will give rise to slightly different evolutionary tracks. This is because although the stellar yields depend solely on $Z_y$, the effective yields ($m_i\tau_{\mathrm{d,s}}/\tau_\star$, i.e. how much of the produced elements end up returning to the ISM) are a function of the system parameters.

Generalizing to allow $x$ to be produced by a non-prompt source with delay time $\tau_\mathrm{p}$, we find for $t_\mathrm{D}=0$ in the exponential star formation rate model that at early times we have
\begin{equation}
\lim_{t\rightarrow0}\frac{Z_x}{Z_y}=C\frac{m_{x,0}}{m_{y,0}}\frac{\tau_\star}{\tau_\mathrm{p}}\frac{Z_y}{m_{y,0}} \Big(1+D\frac{\tau_\star}{\tau_x}\frac{Z_y}{m_{y,0}}\Big),
\end{equation}
where $D=1/3$ for the exponential case and $D=1/4$ for the linear-exponential case, whilst at late times we have
\begin{equation}
\lim_{t\rightarrow\infty}\frac{Z_x}{Z_y}=\frac{m_{x,0}}{m_{y,0}}\frac{\tau_\mathrm{p,s}}{\tau_\mathrm{p}}\Big(1+\frac{\tau_\star}{\tau_x}\frac{Z_y}{m_{y,0}}\Big),
\end{equation}
provided $\tau_\mathrm{p,s}>0$ and $\tau_\mathrm{d,s}>0$. Equating the early- and late-time gradients, we find when $\tau_\mathrm{p,s}<C\tau_x$, [X/Y] linearly rises with [Y/H], plateaus and then continues rising at late times. For long delay times, $\tau_\mathrm{p,s}>C\tau_x$, [X/Y] rises with [Y/H] everywhere and for long $\tau_\mathrm{p}$ steepens at large [Y/H].
If gas is removed from the system faster than it is consumed by star formation and outflow, it is physically possible to have $\tau_\mathrm{d,s}<0$. In this case, the abundance ratio $Z_x/Z_y$ grows linearly in time at late times provided $\tau_\mathrm{p,s}>0$ whilst $Z_x$ increases exponentially. This gives rise to a flattening of [X/Y] vs. [Y/H]. However when $\tau_\mathrm{p,s}<0$ (and still  $\tau_\mathrm{d,s}<0$), [X/Y] vs. [Y/H] rises at late times.

\subsection{Combination of stellar channels}
Until now, we have discussed the evolution of mass fractions from individual channels. A full model is built up from a linear combination of solutions representing each channel. For simplicity, we will here consider two channels, Type II and Type Ia supernovae, but we note that in principle AGB production (or more exotic $r$ process channels) can also be included. Each abundance is then computed as
\begin{equation}
    Z_x(t) = Z_x^{\mathrm{II}}(t)+Z_x^{\mathrm{Ia}}(t).
\end{equation}
We will consider three elements: Fe, Mg and Al. Of these, we assume only Fe has contributions from Type Ia and we will assume only Al has metallicity-dependent yields. We set $m_\mathrm{Fe}^\mathrm{II}/Z_{\mathrm{Fe},\odot}=1$ (corresponding approximately to the \citealt{Kroupa2001} initial mass function yields from \citealt{ChieffiLimongi2004}), $m_\mathrm{Fe}^\mathrm{Ia}=1.3m_\mathrm{Fe}^\mathrm{II}$, $\log_{10}(m_\mathrm{Al,0}^\mathrm{II}/Z_\mathrm{Al,\odot})=-0.3$ and $\log_{10}(m_\mathrm{Mg,0}^\mathrm{II}/Z_\mathrm{Mg,\odot})=+0.35$ \citep[see Table~\ref{tab:yields} and][]{Weinberg}. We set the metallicity yield gradient for Al as $g_\mathrm{Al}m_\mathrm{O}=12$. For the Type Ia delay-time distribution, we choose $t_\mathrm{D}=0.15\,\mathrm{Gyr}$ and $\tau_\mathrm{p}=1.5\,\mathrm{Gyr}$ \citep{Weinberg}.

We will consider two example models: one representing a typical in-situ Milky Way thick disc track and one representing a typical Gaia-Sausage Enceladus (GSE) track. Note that our goal here is to give example models rather than find the `best' model for each system. We use the linear-exponential star formation history (see Table~\ref{tab:equation_navigator} for references to the relevant equations) such that each model is specified by three parameters, $\tau_\mathrm{d}$, $\tau_\star$ and $\tau_\mathrm{s}$. We set $\tau_\star=1\,\mathrm{Gyr}$ for both models and set $\tau_\mathrm{d}=0.5\,\mathrm{Gyr}$ ($\eta=1$) and $\tau_\mathrm{s}=2.5\,\mathrm{Gyr}$ for the Milky Way (thick disc) and $\tau_\mathrm{d}=0.05\,\mathrm{Gyr}$ ($\eta=19$) and $\tau_\mathrm{s}=4.5\,\mathrm{Gyr}$ for GSE. Recall that $\eta$ is an effective mass loading factor that includes the effects of instantaneous recycling. These models are illustrative, and it is understood that a more realistic Milky Way model would have to consider a range of populations. We compute both tracks for $5\,\mathrm{Gyr}$ of evolution.

The resulting tracks are shown in Fig.~\ref{fig::example_tracks}. We observe that the higher outflow efficiency, $\eta=\tau_\star/\tau_\mathrm{d}$, of the GSE model has resulted in a lower final metallicity and a lower metallicity $[\alpha/\mathrm{Fe}]$ `knee' than the Milky Way model.
Note the final metallicity is governed solely by the mass-loading factor, $\eta$, whilst the knee location can be moved to lower metallicities by lowering the star formation efficiency (increasing $\tau_\star$).
We see the metallicity dependence of the Type II Al yields produces the approximately quadratic rise of [Al/Fe] with [Fe/H], which turns over when the Type Ia supernovae begin contributing Fe. The [Mg/Fe] vs. [Al/Fe] \citep[{an analogue of the [Mg/Mn] vs. [Al/Fe] diagrams analysed by e.g.}][although note that Mn has strongly metallicity-dependent yields whilst Fe does not, so the analogy is not perfect]{Das2020, Horta2021} evolve horizontally due to the metallicity dependence of Al production until the Type Ia contribution leads to a reduction in both [Mg/Fe] and [Al/Fe] due to the Fe production. Type Ia production begins contributing at a fixed time in a system's evolution which corresponds to a different metallicity depending on the star formation efficiency and mass-loading factor. This causes the differing turn-offs in the [Mg/Fe] vs. [Al/Fe] diagram and the resulting separation of populations. Finally, the [Al/Mg] vs [Mg/H] has the characteristic quadratic rise due to the linear metallicity dependence of the Al yield on Mg abundance and the tracks are only weakly sensitive to the differences in the system parameters.

\subsection{Impact of model parameter variation on [Al/Fe] vs. [Mg/Fe] plane}
\begin{figure*}
    \centering
    \includegraphics[width=\textwidth]{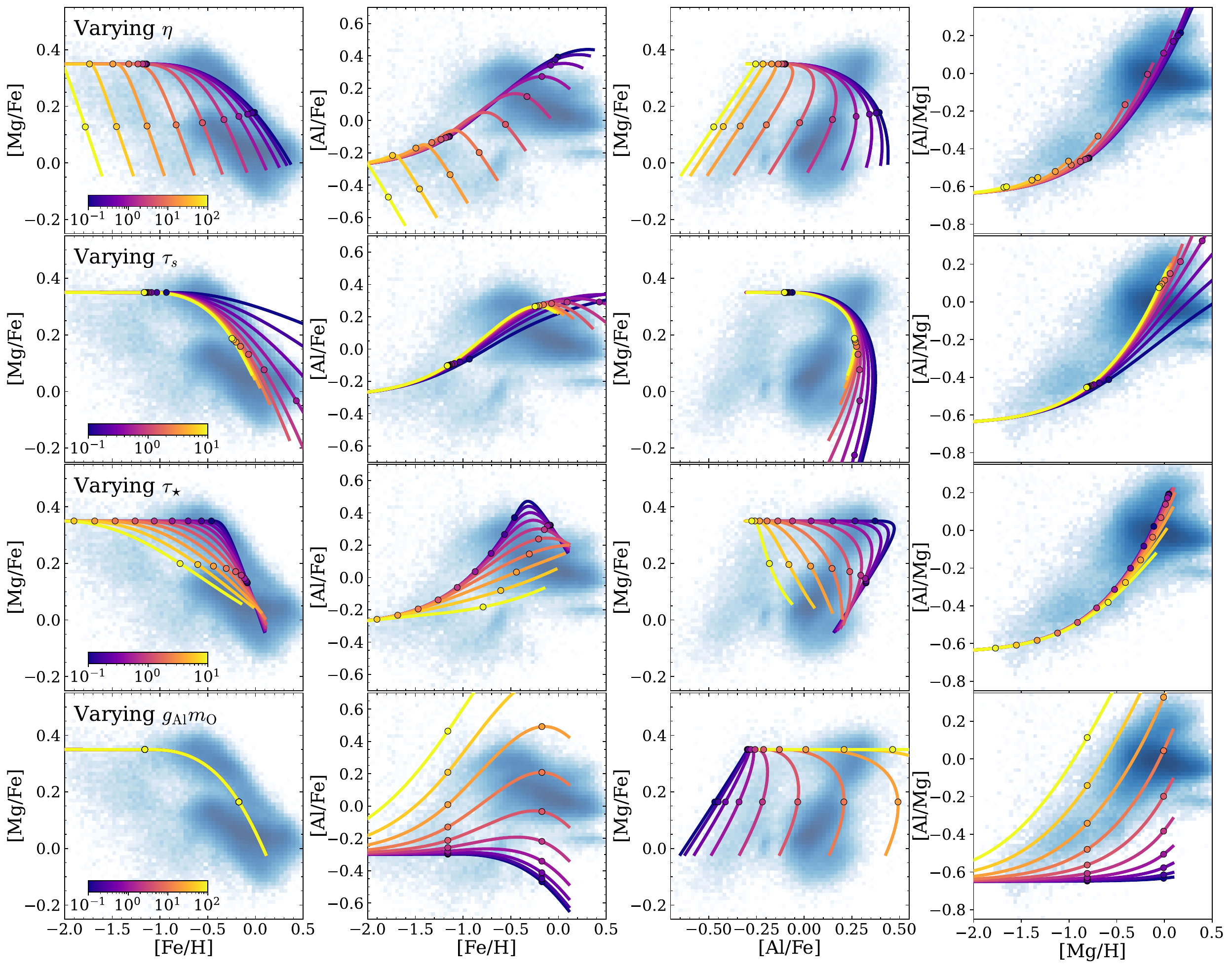}
    \caption{Linear-exponential model variations when changing different parameters. The background blue distribution is APOGEE DR17 stars as shown in Fig.~\ref{fig::example_tracks}. Each row corresponds to varying a different parameter as denoted in the panel: first, mass-loading factor, $\eta$, second, star formation history timescale, $\tau_\mathrm{s}$, third, star formation efficiency timescale, $\tau_\star$ and fourth, the gradient of Type II Al yields with metallicity, $g_\mathrm{Al}$. The colours of the lines correspond to the colour bar in the leftmost panels. The default parameters chosen to emulate the Milky Way thick disc are $\eta=1$, $\tau_\star=1\,\mathrm{Gyr}$ (so $\tau_\mathrm{d}=0.5\,\mathrm{Gyr}$),
    $\tau_\mathrm{s}=2.5\,\mathrm{Gyr}$, and $g_\mathrm{Al}m_\mathrm{O}=12$ (see Table~\ref{tab:yields}). The tracks are all computed for $5\,\mathrm{Gyr}$ and the dots show $t=0.15\,\mathrm{Gyr}$ (when the first Type Ia supernovae start contributing) and $t=2\,\mathrm{Gyr}$.}
    \label{fig:param_grid}
\end{figure*}

In Fig.~\ref{fig:param_grid}, we illustrate further the behaviour of the models in the same chemical abundance planes as Fig.~\ref{fig::example_tracks} by exploring the model dependence on four key parameters: (i) the mass-loading factor, $\eta=\tau_\star/\tau_\mathrm{d}-1$, (ii) the star formation history timescale $\tau_\mathrm{s}$, (iii) the star formation efficiency timescale $\tau_\star$ (at fixed $\eta$ so $\tau_\mathrm{d}$ is adjusted) and (iv) the gradient of the Type II Al yields with metallicity, $g_\mathrm{Al}$. The default parameters are those of the Milky Way model from Fig.~\ref{fig::example_tracks}. It is apparent that the chemical evolution tracks permitted by the models encompass a wide range of behaviour, broadly consistent with the diversity observed in real systems \citep[e.g. figure 2 from][]{BelokurovKravtsov2022}.

As discussed by \cite{Fernandes2023}, the different tracks in the [Al/Fe] vs. [Mg/Fe] panels of Figure~\ref{fig:param_grid} can be characterised by two numbers: [Al/Fe] at the point [Mg/Fe] begins decreasing and the subsequent gradient. We will consider two points: ([Al/Fe], [Mg/Fe])$_0$, the turn-off point, and ([Al/Fe], [Mg/Fe])$_\infty$, the asymptotic abundance ratios at long times, $t\rightarrow\infty$. From equation~\eqref{eqn::stitch}, the location of the turn-off is approximately
\begin{equation}
 \begin{split}
       [\mathrm{Al/Fe}]_0 &\approx \log_{10}\frac{m^\mathrm{II}_{\mathrm{Al},0}/Z_{\mathrm{Al},\odot}}{m^\mathrm{II}_{\mathrm{Fe},0}/Z_{\mathrm{Fe},\odot}}+\log_{10}\Big(1+\frac{Ct_\mathrm{Ia}/\tau_\mathrm{Al}}{1+Ct_\mathrm{Ia}/\tau_\mathrm{d,s}}\Big),\\
       [\mathrm{Mg/Fe}]_0 &= \log_{10}\frac{m^\mathrm{II}_{\mathrm{Mg},0}/Z_{\mathrm{Mg},\odot}}{m^\mathrm{II}_{\mathrm{Fe},0}/Z_{\mathrm{Fe},\odot}},
 \end{split}
 \label{eqn::zero_positions}
\end{equation}
where $C=1/3$ for the linear-exponential case we are considering and $t_\mathrm{Ia}\approx150\,\mathrm{Myr}$ is the minimum delay-time of Type Ia supernovae. We use $\tau_\mathrm{p}$ to refer to the exponential delay time of the Type Ia supernovae. Expressions for ([Al/Fe], [Mg/Fe])$_\infty$ depend on the relative values of $\tau_\mathrm{s}$, $\tau_\mathrm{d}$ and $\tau_\mathrm{p}$. For the typical case of $\tau_{\mathrm{d},\mathrm{s}}>0$ and $\tau_\mathrm{p,s}>0$ (the star formation timescale is significantly longer than the depletion time and the Type Ia delay time distribution timescale), we approach equilbrium values given by (equation~\eqref{eqn::exp_late})
\begin{equation}
\begin{split}
    [\mathrm{Al/Fe}]_\infty &= [\mathrm{Mg/Fe}]_\infty +\log_{10}\frac{m^\mathrm{II}_{\mathrm{Al},0}/Z_{\mathrm{Al},\odot}}{m^\mathrm{II}_{\mathrm{Mg},0}/Z_{\mathrm{Mg},\odot}}+\log_{10}\Big(1+\frac{\tau_\mathrm{d,s}}{\tau_\mathrm{Al}}\Big),\\
[\mathrm{Mg/Fe}]_\infty&=[\mathrm{Mg/Fe}]_0-\log_{10}\Big(1+\frac{m^\mathrm{Ia}_{\mathrm{Fe},0}}{m^\mathrm{II}_{\mathrm{Fe},0}}\mathrm{e}^{t_\mathrm{D}/\tau_\mathrm{s}}\frac{\tau_{\mathrm{p,s}}}{\tau_\mathrm{p}}\Big).
\end{split}
 \label{eqn::inf_positions}
\end{equation}
If $\tau_{\mathrm{d},\mathrm{s}}<0$ and $\tau_{\mathrm{p},\mathrm{s}}>0$ (long depletion time), $[\mathrm{Mg/Fe}]$ tends to a finite limit of
\begin{equation}
[\mathrm{Mg/Fe}]_\infty=[\mathrm{Mg/Fe}]_0-\log_{10}\Big(1+\frac{m^\mathrm{Ia}_{\mathrm{Fe},0}}{m^\mathrm{II}_{\mathrm{Fe},0}}\mathrm{e}^{t_\mathrm{D}/\tau_\mathrm{p}}\frac{\tau_{\mathrm{d,p}}}{\tau_\mathrm{p}}\Big),
\end{equation}
(note the small difference in the final timescale ratio with respect to equation~\eqref{eqn::inf_positions}) whilst $[\mathrm{Al/Fe}]$ increases indefinitely. For $\tau_{\mathrm{d},\mathrm{s}}<0$ and $\tau_{\mathrm{p},\mathrm{s}}<0$ (very short star formation timescale), both [Mg/Fe]$_\infty$ and [Al/Fe]$_\infty$ tend to $-\infty$ as the iron returned from Type Ia enriches the ISM long after stars have stopped being formed in any significant quantity. This is somewhat misleading as the density of the population at these very late times will be vanishingly small, so a clearer comparison would perhaps be with the abundance ratios at e.g. $t=3\tau_\mathrm{s}$. Such calculations are possible with the presented models, but we do not give full expressions.

With these analytic expressions, we can understand quantitatively the structure of the diagrams in Fig.~\ref{fig:param_grid}.

\emph{Varying $\eta$}: We see higher $\eta$ corresponds to lower metallicity $\alpha$ knees and lower equilibrium abundances \citep{Andrews2017, Weinberg}. Correspondingly, this causes the [Al/Fe] vs. [Fe/H] tracks to turn over at lower metallicities but the gradient beyond the turn-over flattens with decreasing $\eta$ as more Al is produced at the higher metallicities reached. The [Mg/Fe]-[Al/Fe] plane shows that lower $\eta$ drives [Al/Fe] higher before the Type Ia contribution begins (due to the $Ct_\mathrm{Ia}/\tau_{\mathrm{d,s}}=(Ct_\mathrm{Ia}/\tau_\star)(1+\eta-\tau_\star/\tau_\mathrm{s})$ in the denominator of equation~\eqref{eqn::zero_positions}) and the trend beyond this tends to 1:1 at high $\eta$, where the metallicity dependence of Al production is insignificant, to near vertical as the increased Al production with increasing metallicity offsets the increased iron production from Type Ia supernovae. Mathematically, this is driven by equation~\eqref{eqn::inf_positions} where we see the final bracket in [Al/Fe]$_\infty$ decreases with increasing $\eta$ and large $\eta$ means the additional metallicity-dependent terms in [Al/Fe]$_0$ and [Al/Fe]$_\infty$ tend to zero and so [Al/Fe] behaves identically to [Mg/Fe]. Finally, the [Al/Mg] vs [Mg/H] is only weakly a function of $\eta$ as discussed in Section~\ref{sub:abundance_ratios_from_individual_channels}.

\emph{Varying $\tau_\mathrm{s}$}: the star formation timescale has a rather weak effect on the evolutionary tracks, largely because the key quantity in the equations is $\tau_\star/\tsfh$ so when $\tsfh\gg\tau_\star$ the mass loading factor, $\eta$, has a more important effect. Here we have chosen a low default value of $\eta$ so this already weak dependence will only get weaker in lower mass systems with lower $\eta$. However, note that $\tsfh$ also affects the density of star formation along each track. We see from equations~\eqref{eqn::zero_positions} and~\eqref{eqn::inf_positions} that [Al/Fe]$_0$ and [Al/Fe]$_\infty$ decrease with decreasing $\tau_\mathrm{d,s}$ and hence increasing $\tsfh$. When $\tau_\mathrm{s}<\tau_\mathrm{d}$, on timescales $t>\tau_{\mathrm{s,d}}$, the system will have negative inflow i.e., the gas supply is decreasing more rapidly than it is being consumed by star formation and outflow alone. We see from Fig.~\ref{fig:param_grid} that these negative inflow models have much stronger sensitivity to $\tsfh$ as large quantities of metals are returned to a near-depleted ISM, thus vastly increasing the metallicity although the star formation from this high metallicity material will be very low. Whilst physically possible \citep[e.g.][mentions a ram-pressure-stripped system may have this property]{Weinberg}, these cases are not of broad applicability.

\emph{Varying $\tau_\star$}: varying the star formation efficiency has a significant effect on the models. Fundamentally higher star formation efficiency (lower $\tau_\star$) means more metals returned per unit gas mass and the models are driven to higher gas mass at earlier epochs (e.g. equation~\eqref{eqn::lexp_early_td0}). However, at late times the tracks all converge to a point independent of $\tau_\star$ \citep{Andrews2017}. We see from equation~\eqref{eqn::zero_positions} that $\tau_\star$ dependence in [Al/Fe]$_0$ arises from both $\tau_\mathrm{Al}$ and $\tau_\mathrm{d,s}$ where $\tau_\mathrm{Al}=\tau_\star/g_\mathrm{Al}m_\mathrm{O}$ appears the more significant term for the early evolution. Higher $\tau_\star$ leads to higher gas mass into which the products are diluted so lower metallicities at early times and hence lower [Al/Fe]$_0$. Increasing $\tau_\star$ also leads to longer depletion times, hence longer $\tau_\mathrm{d,s}$ and slower approach towards equilibrium. For higher $\tau_\star$, when the Type Ia begin producing Fe, the enhanced Al from the Type II metallicity dependence is being offset by the increased Fe production from Type Ia, leading to a flatter track in [Al/Fe]. We see in the [Al/Fe] vs. [Mg/Fe] diagram that high $\tau_\star$ leads to negative gradients after the turn-down and low values produce positive gradients. Inspecting equations~\eqref{eqn::zero_positions} and~\eqref{eqn::inf_positions}, the critical value of $\tau_\star$ where [Al/Fe]$_0=$[Al/Fe]$_\infty$ is
\begin{equation}
    \tau_{\star,\mathrm{crit}}=\frac{Ct_\mathrm{Ia}(1+g_\mathrm{Al}m_\mathrm{O}+\eta)(1+\eta)q}{g_\mathrm{Al}m_\mathrm{O}-(1+\eta)q},
\end{equation}
where $q=(m^\mathrm{Ia}_{\mathrm{Fe},0}/m^\mathrm{II}_{\mathrm{Fe},0})\mathrm{e}^{t_\mathrm{D}/\tau_\mathrm{s}}\tau_{\mathrm{p,s}}/\tau_\mathrm{p}\approx3.5$ for the considered example. $\tau_{\star,\mathrm{crit}}$ is principally a function of the mass loading factor, $\eta$, with a weaker dependence on the specifics of the star formation law through $C$ and $q$. For the example shown in Fig.~\ref{fig:param_grid}, $\eta=1$, $C=1/3$, $t_\mathrm{Ia}=0.15\,\mathrm{Gyr}$ and $g_\mathrm{Al}m_\mathrm{O}=12$, so $\tau_{\star,\mathrm{crit}}\approx1\,\mathrm{Gyr}$, agreeing well with those curves that are approximately vertical after the turn-down. If $\tau_\star<\tau_{\star,\mathrm{crit}}$ the tracks have a positive gradient whilst for $\tau_\star>\tau_{\star,\mathrm{crit}}$ the gradient is negative. Note for $\eta>\eta_\mathrm{crit}=g_\mathrm{Al}m_\mathrm{O}/q-1\approx2.5$, the gradient will always be positive, although this value depends on the choice of $\tau_\mathrm{s}$ ($q$ is anticorrelated with $\tau_\mathrm{s}$) and the Type Ia delay time distribution. This equates to $\alpha$-knee locations $[\mathrm{Fe}/\mathrm{H}]\lesssim-1\,\mathrm{dex}$.

\emph{Varying $g_\mathrm{Al}$}: the yield gradient $g_\mathrm{Al}$ is not a system parameter but rather is related to the yields. Therefore, inspecting the model behaviour when varying it allows us to consider what features in the data constrain this somewhat uncertain parameter (e.g. see Fig.~\ref{fig:metal_production}) but also how different elements with different yield gradients with metallicity behave. Increasing $g_\mathrm{Al}$ naturally increases [Al/Fe] at fixed [Fe/H] and we see in the [Al/Fe] vs. [Mg/Fe] plane, its variations are somewhat degenerate with the mass loading factor $\eta$. We observe that of the model parameters investigated $g_\mathrm{Al}$ variation gives rise to the strongest variation in [Al/Mg] vs. [Mg/H] meaning this space puts strong constraints on the metallicity dependence of the yields rather than anything related to an individual system. However, note that system-to-system variations in the initial mass function could lead to Type II yield variations, as production from low- versus high-mass stars varies.

\cite{Fernandes2023} discuss how the direction of increasing metallicity of stellar populations in the [Al/Fe] vs. [Mg/Mn] plane reflects variations in the (early) star formation history of their progenitor system. For example, the Milky Way disc population evolves from high [Al/Fe], high [Mg/Mn] at low metallicity to low [Al/Fe], low [Mg/Mn] at high metallicity whilst the LMC population evolves from low [Al/Fe], high [Mg/Mn] at low metallicity to high [Al/Fe], low [Mg/Mn] at high metallicity. We can assess the claim that this is driven by star formation timescales by identifying the distributions of stars from a single system with the [Al/Fe] vs. [Mg/Fe] tracks from one of our single-zone chemical evolution models, again recognising that Fe is an imperfect proxy for Mn due to its metallicity-dependent yields (Fig.~\ref{fig:metal_production}). Fig.~\ref{fig:param_grid} shows that varying the star formation history distribution ($\tau_\mathrm{s}$) has a weak effect on the slope of the chemical evolution tracks in the [Al/Fe] vs. [Mg/Fe] plane and the larger effect on the slope is due to varying the \emph{star formation efficiency} and to a lesser extent the mass loading factor. In particular, our solutions only admit tracks that run from low [Al/Fe], high [Mg/Fe] to high [Al/Fe], low [Mg/Fe] when the mass loading factor is low, $\eta\lesssim2.5$ \emph{and} the star formation efficiency is low. These correspond to systems like the LMC with intermediate stellar mass and hence low mass loading factors, but also crucially low star formation efficiency. ponse{As shown in figure 5 of \cite{Fernandes2023}, dwarf spheroidal systems like Fornax and Sculptor run from high [Al/Fe], high [Mg/Mn] to low [Al/Fe], low [Mg/Mn], which we suggest is driven by their low mass and hence high mass loading factor above the critical threshold $\eta_\mathrm{crit}=g_\mathrm{Al}m_\mathrm{O}/q-1$. The star formation rate timescale $\tau_\mathrm{s}$ and the star formation timescale $\tau_\star$ can be physically correlated via the inflow rate, and for galaxies with negligible inflows, the timescales are equal. However, our analysis has demonstrated that the physical driver for different directions of increasing metallicity in the [Al/Fe]-[Mg/Mn] plane for different populations is likely linked to star formation efficiency rather than the star formation history, with the caveat that we have not fully modelled the potential metallicity dependence of the Mn yields.

\cite{Hasselquist2021} and \cite{BelokurovKravtsov2022} found that in-situ and ex-situ (dSph or GSE) stars follow different tracks in [Al/Mg] vs. [Mg/H] \citep[an offset of $\sim0.2\,\mathrm{dex}$ between in-situ and GSE stars at fixed $\mathrm{[Mg/H]}$ was observed by][]{BelokurovKravtsov2022}. Figure~\ref{fig:param_grid} shows the variation of [Al/Mg] vs. [Mg/H] tracks with varying model parameters is very weak. The fact that there is any variation at all is somewhat surprising as we have assumed Mg and Al are prompt (returned as soon as the stars have formed) and the Al yield is then a function of Mg (as Mg behaves identically to O). However, equation~\eqref{eqn::z_x_z_y_with_z_y} shows that varying $1+\eta-\tau_\star/\tsfh$ \emph{can} produce variations of [Al/Mg] at fixed [Mg/H] because the models depend on the effective yield $m_i\tau_{\mathrm{d,s}}/\tau_\star$. Figs.~\ref{fig::example_tracks} and~\ref{fig:param_grid} show that in practice such variations are small, and this is particularly true if we were to further fix the models to reproduce a given $\alpha$ knee and late time metallicity limit.
As our metallicity-dependent yield model has not introduced any new system parameters, fixing the system parameters using other chemical spaces necessarily prescribes the [Al/Mg] vs. [Mg/H] evolution. Therefore, any further variation in [Al/Mg] (such as that seen by \cite{BelokurovKravtsov2022}) hints perhaps at the need for model generalizations such as the introduction of other stellar channels (note in all current AGB models Al production is at the percent level compared to the production in Type II supernovae so this is unlikely to change things) or a different initial mass function, for example.

In summary, our models have given a mathematical description of the expected behaviour of different stellar populations in the [Al/Fe] vs. [Mg/Fe] plane described in the Introduction: the primary factor for the locus of a population in this space is the mass-loading parameter, $\eta$. Lower $\eta$ means higher metallicities are reached so higher characteristic [Al/Fe] abundances are reached before Type Ia supernovae begin contributing significant Fe. Here we have elucidated for the first time the secondary role that the star formation efficiency, $\tau_\star^{-1}$, plays with low star formation efficiencies leading to decreasing [Mg/Fe] but increasing [Al/Fe] at late times, such as seen in the LMC.

\section{
Generalizations and limitations
}\label{sec::generalizations}
We have demonstrated the applicability of our models to data from spectroscopic surveys despite the assumptions required to make the solutions analytic. We here discuss these limitations further and sketch some possible generalizations of our models to incorporate more complex behaviour.

\emph{Yield dependence on a delayed element}: We have chosen to simplify the presented solutions by making the effective metallicity dependence a dependence on the oxygen mass fraction (or more generally any purely prompt element). The evolution of oxygen is simple, so the resulting metallicity-dependent pieces of the secondary element mass fraction are (relatively) simple. As described in the introduction, the physics driving metallicity-dependent yields is more complicated than this. The s-process begins with seed iron nuclei and requires free neutrons that arise in $^{13}$C pockets \citep{KarakasLattanzio2014}. However, the dynamic range of $N_\mathrm{O}/N_\mathrm{Fe}$ is a factor of $2-3$ whilst the dynamic range of $N_\mathrm{O}/N_\mathrm{H}$ or $N_\mathrm{Fe}/N_\mathrm{H}$ is much larger ($100-1000$). Therefore, even if the physical dependence is on iron, tying the metallicity dependence of the yields to oxygen is likely to be a reasonable approximation. Nonetheless, in our formulation, we could have made the metallicity dependence a function of the iron abundance (or even total heavy element abundance). Iron is produced in both Type II and Type Ia supernovae. For the Type-II-dependent part, we would obtain identical equations to the oxygen-dependent cases already presented. With a dependence upon an element produced in Type Ia supernovae, we have to consider the additional delay time from this channel. In general, this delay could be different from the delay in the production of element $x$, which could arise from neutron star mergers or AGB stars, for example. In the case where the delayed secondary production of an element $x$ depends on the delayed primary production of an element $y$ through the same stellar channel with equal delay times e.g. possibly nitrogen production dependent on carbon production, we would write (in the case $t_\mathrm{D}=0$)
\begin{equation}
U_x(s) = m_{x,0}\Gamma(s)\frac{1}{s+\tau_\mathrm{d}^{-1}}\frac{\tau_\mathrm{p}^{-1}}{s+\tau_\mathrm{p}^{-1}}\frac{\tau_x^{-1}}{s+\tau_\mathrm{d}^{-1}}\frac{\tau_\mathrm{p}^{-1}}{s+\tau_\mathrm{p}^{-1}},
\end{equation}
so
\begin{equation}
Z_x(t) = \frac{1}{\tau_x\tau_\mathrm{p}^2}\frac{\partial}{\partial\tau_\mathrm{p}^{-1}}\tau_\mathrm{p}\frac{\partial}{\partial\tau_\mathrm{d}^{-1}}Z_y(t).
\label{eqn::simple_metal_second}
\end{equation}
We, therefore, assert that the solutions in these general cases for the choices of $\Gamma(s)$ in Table~\ref{table:sfr_laplace} are all analytically tractable. However, they lack the simplicity of the already presented solutions. It is possible they are useful from a more computational standpoint.

\emph{Yield dependence on a `secondary element'}: It has been suggested that the production of some elements depends upon the mass fraction of an element produced predominantly through secondary processes (sometimes these are dubbed double secondary or tertiary). For example, carbon might have a secondary component \cite[possibly through stellar winds,][]{Ma2025} although, as we have seen, delayed production is degenerate with secondary production \citep{Chiappini2003}. \cite{Henry2000} presented analytic solutions for the production of nitrogen assuming a linear dependence on the carbon mass fraction, which itself had a linear dependence on the metallicity. If we consider a tertiary yield that depends on the abundance of an element $y$ produced through secondary processes in Type II supernovae ($S_x(t)=m_{x,0}(1+g_x'Z_y(t))\sfr(t)$, a slightly artificial example that serves to illustrate the generalization), equation~\eqref{eqn::x_neq_y} would read
\begin{equation}
\begin{split}
U_x(s) &= m_{x,0}\Gamma(s)\frac{\mathrm{e}^{-t_\mathrm{D}s}}{\tau_\mathrm{p}}\frac{1}{s+\tau_\mathrm{d}^{-1}}\frac{1}{s+\tau_\mathrm{p}^{-1}}\Big(1+\frac{\tau_y^{-1}{\tau'_x}^{-1}}{(s+\tau_\mathrm{d}^{-1})^2}\Big),\\
&= \Big(1+\frac{1}{\tau_x'\tau_y}\frac{\partial^2}{\partial(\tau_\mathrm{d}^{-1})^2}\Big)U^0_x(s),
\end{split}
\end{equation}
where $\tau_x'=\tau_\star/(g_x'm_{y,0})$. This demonstrates that tertiary elements can also be incorporated into our framework.

\emph{General metallicity dependence of the yields}: Arbitrary dependence of the yields on metallicity may be required for the most accurate modelling. Naturally, arbitrary dependence will not yield analytic solutions so these cases would have to be solved numerically in the general case. Our formalism suggests a generic methodology for solving the chemical evolution equations with more arbitrary metallicity dependence via the Laplace transform, but typically numerical inverse Laplace transforms are unstable \citep{epstein2008}. However, based on the example of saturating the linear growth, a piecewise linear function for the metallicity dependence of the yields is also tractable using our methods, and it is likely this is sufficient for most applications.

\emph{More complicated star formation histories}: We have presented solutions for two choices of star formation history, exponential and linear-exponential (with constant star formation as the limit of the exponential when $\tsfh\rightarrow\infty$). It is likely there are other choices of star formation history that yield analytic solutions e.g. Gaussians.
However, our models are immediately generalizable to more realistic use cases by considering linear combinations of the presented solutions. \cite{Weinberg2023} advocate using a linear combination of two exponential star formation rates $\sfr\propto(1-\mathrm{e}^{-t/\tau_1})\mathrm{e}^{-t/\tau_2}$ as a balance between simplicity and physical realism. For $\tau_1<\tau_2$, this model has an early rise on a timescale $\tau_1$ and a late time fall on timescale $\tau_2$. Therefore, it approximates a linear exponential star formation rate, but with more control over the rise and peak location. In Table~\ref{table:sfr_laplace}, we give the Laplace transform of this star formation rate (a linear combination of the Laplace transform of the exponential). The solution for $M_x(t)$ can then be constructed by a difference of the exponential solutions in Section~\ref{sub:exponential_star_formation_rate} with timescales $\tau_2$ and $\tau_\mathrm{h}^{-1}\equiv\tau_1^{-1}+\tau_2^{-1}$ \citep[see the appendix in][for related discussion]{Weinberg2023}.

One other interesting route to constructing a general star formation history, $f(t)$, is projection onto a Laguerre polynomial ($L_n(x)$) basis as
\begin{equation}
    \sfr(t) = \sum_n c_n L_n(t/\tsfh) \mathrm{e}^{-t/\tsfh},
\end{equation}
where the coefficients are
\begin{equation}
c_m=\tsfh^{-1}\int_0^{\infty} \mathrm{d}t' L_m(t'/\tsfh) \sfr(t'),
\end{equation}
for some choice of $\tsfh$. This is practical as (i) the representation is unique and (ii) the monomial parts of the Laguerre polynomials give rise to star formation histories of the form $\sfr^n(t)=\mathrm{e}^{-t/\tsfh}(t/\tsfh)^n/n!$, the $n$th derivative of the exponential star formation history. The solution for the general star formation history can therefore be approximately computed analytically via derivatives of the solutions presented in this paper (in a similar way to equation~\eqref{eqn::exp_to_linearexp}). For exponential delay-time distributions, recurrence relations exist to make computation easier. For example, if $\tau_\mathrm{p}=0$ $M_x^n(t) = \tau_{\mathrm{d,s}}(\sfr^n(t)-M_x^{n-1}(t))$ where $M_x^n(t)$ is the solution for the mass of element $x$ for star formation history $\sfr^n(t)$. Metallicity dependent solutions can be derived by differentiation i.e. equation~\eqref{eqn::simple_metal}. From a modelling perspective, $c_n$ can be considered as free parameters to model arbitrary star formation histories and rapidly generate the corresponding chemical evolution history.

\emph{More complicated star formation efficiency and outflow histories}: We have seen how the assumptions in the modelling can be relaxed and generalized to give access to more general (but mathematically more complex) solutions. Two fundamental limitations of our framework are the constant star formation efficiency (linear Kennicutt-Schmidt law) and the constant mass-loading factor. Relaxing these would likely require fully numerical solutions although probably linear time dependence is tractable \citep{Weinberg}. It is however possible to generalize the models to make these variables piecewise constant functions \citep[see e.g.][]{Weinberg}. Denoting the solution for mass of element $x$ using a depletion time $\tau_\mathrm{d}$ as $M_x(t; \tau_\mathrm{d})$, we consider a depletion time $\tau_\mathrm{d}=\tau_{\mathrm{d},0}$ for $t\leq t_0$ and $\tau_\mathrm{d}=\tau_{\mathrm{d},1}$ for $t>t_0$. The solution for $t<t_0$ is $M_x(t; \tau_{\mathrm{d},0})$ whilst for $t>t_0$ we find
\begin{equation}
    M_x(t>t_0) = M_x(t;\tau_{\mathrm{d},1})-\mathrm{e}^{-(t-t_0)/\tau_{\mathrm{d},1}}\Big(M_x(t_0;\tau_{\mathrm{d},1})-M_x(t_0;\tau_{\mathrm{d},0})\Big).
\end{equation}
$Z_x$ is then found by dividing $M_x$ by $M=\tau_\star(t)\sfr(t)$.
Further splits can naturally be included and the metallicity-dependent yield solutions can be derived by differentiation (equation~\eqref{eqn::simple_metal}).

\emph{Enriched infall}: We have assumed that the inflowing gas is pristine. Enriched infall from gas with metallicity $Z_\mathrm{inf}$ modifies equation~\eqref{eqn::main} as $\mdot_x +\tdepi M_x =\mathcal{G}_x+Z_{x,\mathrm{inf}}\dot{M}_\mathrm{inf}$ where $\dot{M}_\mathrm{inf}$ is the infall rate given by $\dot{M}=-\sfr(1+\eta)+\dot{M}_\mathrm{inf}$. $\dot{M}=\tau_\star\ddot{M_\star}$ by the linear Kennicutt-Schmidt law, so for the exponential star formation history, $\dot{M}_\mathrm{inf}=(\tau_\star/\tau_{\mathrm{d},\mathrm{s}})\sfr$. This means enriched infall behaves identically to a prompt source term with effective yield $m$ (mass per unit star formation) of $m=Z_{x,\mathrm{inf}}\tau_\star/\tau_{\mathrm{d},\mathrm{s}}$ \citep{Weinberg}. If $\mathcal{G}_x$ is composed solely of prompt metallicity-independent contributions, this is equivalent to enhancing the yield by $m_{x,0}\leftarrow m_{x,0}(1 + Z_{x,\mathrm{inf}}/Z^0_{x,\mathrm{eq}})$. Further results (e.g. linear-exponential star formation history, metallicity dependence of the yields, warm interstellar medium phase) can be derived from this result from differentiation or linear superposition (equations~\eqref{eqn::simple_metal}, ~\eqref{eqn::exp_to_linearexp} and~\eqref{eqn:warm_linearsup}).

\section{Conclusions}\label{sec::conclusions}
We have presented a series of analytic chemical evolution models describing the enrichment of elements with some metallicity dependence to their stellar yields. Elements with metallicity dependent yields, most notably aluminium, are some of the most interesting due to their use in separating out stellar populations from different galactic environments.

\subsection{Summary of results}
Our results are as follows:
\begin{enumerate}[(i)]
    \item We have demonstrated how under the assumption of a linear Kennicutt-Schmidt type law and constant mass-loading factor, single-zone models with linear dependence of the yields on a prompt element (e.g. O) can be derived analytically from the corresponding metallicity-independent result. This includes cases where element production is delayed through some delay-time distribution. The metallicity dependence mathematically behaves like a system-dependent delay time approximately equal to the depletion time. The depletion time governs the timescale on which primary production increases and hence the timescale on which the metallicity dependence of the yields becomes important.
    \item We have given full analytic solutions for the cases where the star formation history is constant, exponential or linear exponential and the delay-time distribution is given by an exponential. Appendix~\ref{appendix::alternative_functional_forms_for_the_delay_time_distribution} gives further solutions for Gaussian or $1/t$ power law delay-time distributions.
    \item We have discussed generalizations to these solutions via linear superposition (e.g. incorporating multiple stellar channels, incorporating a warm phase that delays the return of products to the star-forming interstellar medium),  differentiation with respect to the model parameters (e.g. to construct fully general star formation histories through a Laguerre polynomial series or to incorporate more complex yield dependence on elements produced through non-primary processes) or by introducing a saturation to the linear metallicity dependence of the yields.
    \item We have presented comparisons of our solutions with APOGEE DR17 data including the construction of models of the Milky Way thick disc and the accreted \emph{Gaia-Sausage Enceladus} galaxy. The simple models are a reasonable match to the data.
    \item A full parameter investigation of the [Al/Fe] vs. [Mg/Fe] plane, an analogue of the [Al/Fe] vs. [Mg/Mn] plane, that has proven useful for accreted vs. in-situ separation, reveals that populations initially increase [Al/Fe] at fixed [Mg/Fe] due to metallicity dependence of the Al yields before `turning off' at the [Al/Fe] reached by the time the Type Ia supernovae start contributing. This critical value is typically higher for in-situ components than for accreted components and our models have demonstrated the value is governed by the mass-loading factor, star formation efficiency and the yields gradient (the last of which is system-independent), and only quite weakly on the star formation timescale. The populations then primarily decrease both [Al/Fe] and [Mg/Fe] so populations move to the lower left in the diagram, although low star formation efficiencies can cause populations to move towards the lower right as [Al/Fe] continues increasing whilst [Mg/Fe] declines. Our models then demonstrate that in-situ and accreted environments with high star formation efficiency (e.g. Milky Way or GSE) will produce stellar distributions that have a positive correlation in the [Al/Fe] vs. [Mg/Fe] plane whilst environments with low star formation efficiency (e.g. LMC) produce stellar distributions that have a negative correlation.
\end{enumerate}

\subsection{User's Guide}\label{sec::users}
Our work has yielded models that we hope will prove useful in future chemical evolution modelling investigations. This guide summarises the practical steps for implementing the analytic solutions presented in this paper. Readers will also find a similar User's Guide from \cite{Weinberg} useful. In order to implement the models, users should:
\begin{enumerate}[(i)]
    \item Choose star formation history $\sfr(t)$ (exponential, constant or linear-exponential) and the constants $\tau_\star$ and $\eta$ (and hence $\tau_\mathrm{d}=\tau_\star/(1+\eta)$).
    \item For each element, pick the required channels (e.g. Type II supernovae, Type Ia supernovae, etc.) with net yields $m_{x,0}$ (see Table~\ref{tab:yields} for example values) and delay times ($\tau_\mathrm{p}$, $t_\mathrm{D}$).
    \item Compute the primary solution $Z_x^0(t)$ \citep[equations~\eqref{eqn::general_z} or \eqref{eqn:leq_general_z}, see also][]{Weinberg}.
    \item If yields are metallicity-dependent, set the yield gradient $g_x$ (see Table~\ref{tab:yields} for example values) and compute the metallicity-dependent solution $Z_x^1(t)$ (equations~\eqref{eqn:const_sfr_curlyZ} or \eqref{eqn:leq_general_curlyz}) and set $Z_x=Z_x^0+Z_x^1$.
    \item For saturation of the yields at $Z_c$, apply the formulae in equations~\eqref{eqn:trunc_exp} for exponential (constant) star formation history or \eqref{eqn:trunc_lexp} for linear-exponential.
    \item Combine channels linearly: to model more complicated delay-time distributions or rise--fall star formation histories (Table~\ref{table:sfr_laplace}) use linear superpositions; for warm-phase delays, use equations~\eqref{eqn:warm_linearsup} and~\eqref{eqn::warm_linearsup1}.
\end{enumerate}
The most relevant equations are listed in Table~\ref{tab:equation_navigator}. There is a Python implementation of the equations provided at \url{https://github.com/jls713/chem-evol-z-yields}.

\subsection{Outlook}
Our discussion is an initial step towards flexibly modelling stellar elemental abundance distributions. We have not explored the full capabilities of the presented solutions, nor have we attempted to quantitatively fit the models to the data. In particular, we have omitted detailed inspection of linear combinations of exponential delay time distributions which can better approximate more realistic power law distributions (although see Appendix~\ref{appendix::alternative_functional_forms_for_the_delay_time_distribution} for analytic solutions in this case). Furthermore, we have not fully explored the capabilities of the solutions in the case of metallicity-dependent yields through a delayed channel. This case is of interest for (i) nitrogen which has a metallicity-dependent yield from asymptotic branch star production \citep[e.g.][]{Johnson2023} and, as discussed in the Introduction, is a particularly interesting element for both Galactic and extragalactic studies, and (ii) manganese which has a metallicity-dependent Type Ia yield and has received attention as a diagnostic for sub-Chandrasekhar Type Ia supernovae \citep{Seitenzahl2013B, Sanders2021}. Both of these directions merit further future investigation to understand the true potential of the presented solutions.

The coming years will see the continued influx of spectroscopic results from stellar surveys of the Galaxy such as WEAVE, SDSS-V Milky Way Mapper \citep{SDSSDR19} and 4-MOST. The analysis of the increasingly large number of stars with increasingly high-dimensional abundance patterns will require fast, flexible, yet realistic models to constrain both properties of the stellar evolutionary channels and the galactic environments. We hope the solutions presented here will prove useful in this endeavour.

\section*{Acknowledgements}
JLS thanks his grandfather, Geoffrey Jackson, whose copy of \cite{jaeger1962introduction} helped him understand the power of the Laplace transform.
JLS thanks Vasily Belokurov for useful discussions, and David Weinberg for initially motivating this work and his thorough and detailed comments as a referee.
JLS acknowledges the support of the Royal Society  (URF\textbackslash R1\textbackslash191555; URF\textbackslash R\textbackslash 241030) and the kind hospitality of the Institute of Astronomy, University of Cambridge.

Funding for the Sloan Digital Sky Survey V has been provided by the Alfred P. Sloan Foundation, the Heising-Simons Foundation, the National Science Foundation, and the Participating Institutions. SDSS acknowledges support and resources from the Center for High-Performance Computing at the University of Utah. The SDSS web site is \url{www.sdss.org}.

SDSS is managed by the Astrophysical Research Consortium for the Participating Institutions of the SDSS Collaboration, including the Carnegie Institution for Science, Chilean National Time Allocation Committee (CNTAC) ratified researchers, the Gotham Participation Group, Harvard University, Heidelberg University, The Johns Hopkins University, L’Ecole polytechnique f\'{e}d\'{e}rale de Lausanne (EPFL), Leibniz-Institut f\"{u}r Astrophysik Potsdam (AIP), Max-Planck-Institut f\"{u}r Astronomie (MPIA Heidelberg), Max-Planck-Institut f\"{u}r Extraterrestrische Physik (MPE), Nanjing University, National Astronomical Observatories of China (NAOC), New Mexico State University, The Ohio State University, Pennsylvania State University, Smithsonian Astrophysical Observatory, Space Telescope Science Institute (STScI), the Stellar Astrophysics Participation Group, Universidad Nacional Aut\'{o}noma de M\'{e}xico, University of Arizona, University of Colorado Boulder, University of Illinois at Urbana-Champaign, University of Toronto, University of Utah, University of Virginia, Yale University, and Yunnan University.

\section*{Data Availability}
All data used in this paper is in the public domain. The described analytic models are implemented in python, available at \url{https://github.com/jls713/chem-evol-z-yields}.

\bibliographystyle{mnras}
\bibliography{bibliography}

\appendix

\section{When the mass of element \texorpdfstring{$x$}{x} returned depends on the mass fraction of \texorpdfstring{$x$}{x}}\label{appendix::x_eq_y}

In the main body (Section~\ref{section::x_neq_y}), we consider the case where the yield of element $x$ depends on the interstellar medium (ISM) mass fraction of a different element $y$ and focus on solutions where element $y$ is produced by solely prompt channels.

Another interesting initial case to consider is when $x=y$ i.e. the return of the element $x$ depends on the ISM mass fraction of element $x$. The \emph{production} of an element is likely not directly related to the initial mass of this element in the star except in a few cases. However, in the absence of any production, the return of an element is exactly equal to the mass that the star formed with (or a fraction of it if a remnant is formed). Sometimes the distinction is made between a gross yield, the sum of the mass of element $x$ at formation and any additional production of element $x$, and the net yield, solely the newly produced mass. In this recycling case, we are considering a net yield of zero.

We consider production through a single channel with delay time, $\tau_\mathrm{p}$, and we split the recycling of elements into two pieces: `instantaneous' recycling of a mass fraction $r_\mathrm{II}$ through Type II supernovae and delayed recycling of a mass fraction $r_\mathrm{AGB}$ from asymptotic giant branch (AGB) stars. \cite{Weinberg} discusses how for a Kroupa initial mass function, $r_\mathrm{II}=0.2$ and $r_\mathrm{AGB}=0.3$ are reasonable values. The total return fraction is $r=r_\mathrm{II}+r_\mathrm{AGB}$. We use e.g. equation~\eqref{eqn::laplace} to write the Laplace transform for the mass of element $x$ as
\begin{equation}
\begin{split}
U_x(s) = &\frac{\mathrm{e}^{-t_\mathrm{D}s}}{\tau_\mathrm{p}}\frac{m_{x,0}\Gamma(s)}{(s+\tau_\mathrm{dep}^{-1})(s+\tau_\mathrm{p}^{-1})}\\&+\frac{r_\mathrm{II}}{\tau_\star}\frac{U_x(s)}{s+\tau_\mathrm{dep}^{-1}}+\frac{r_\mathrm{AGB}}{\tau_\star}\frac{U_x(s)}{(s+\tau_\mathrm{dep}^{-1})(s+\tau_\mathrm{r}^{-1})\tau_\mathrm{r}}.
\end{split}
\label{eqn::recycling1}
\end{equation}
The first term is the production, the second term is the mass recycling from Type II supernovae and the third term is the mass recycling from AGB stars. Unlike in the main body, $\tau_\mathrm{dep}=\tau_\star/(1+\eta)$ here uses the true mass-loading factor $\eta$ without any implicit correction for recycling. Note that the final two terms are proportional to the Laplace transform of the mass of element $x$ as the quantity of mass recycled depends on the mass present when the stars are formed.
Here $\tau_\mathrm{r}$ is a characteristic recycling timescale. If we consider an initial mass function, $\mathrm{d}N/\mathrm{d}m\propto m^{-2.3}$ and the stellar lifetime is approximately $\tau \propto m^{-2.5}$, we find the number of stars of mass $m(\tau)$ dying per unit time after time $\tau$ is $\mathrm{d}N_\mathrm{dying}/\mathrm{d}\tau\propto m^{1.2}(\tau)$. If each of the stars returns a fixed fraction of their mass, the total mass returned per unit time is $\mathcal{D}(\tau)\propto \mathrm{d}M_\mathrm{returned}/\mathrm{d}\tau \propto m^{2.2}(\tau)\propto \tau^{-0.9}$. Equation~\eqref{eqn::recycling1} approximates this power law delay-time distribution as a single exponential for which $\tau_\mathrm{r}=3\,\mathrm{Gyr}$ is a reasonable choice but more precise modelling would have to consider sums of exponential delay-time distributions \citep[akin to][]{Weinberg, Sanders2021} or possibly approximate the power-law as $\tau^{-1}$ for which solutions in terms of exponential integral functions may exist (Appendix~\ref{appendix::alternative_functional_forms_for_the_delay_time_distribution}). Furthermore, we have not introduced a minimum delay time to AGB recycling. An exponential delay time distribution already underpredicts early time recycling, so this is not a concern.

We can rearrange equation~\eqref{eqn::recycling1} to give
\begin{equation}
U_x(s)\Big[(s+\tau_\mathrm{d}^{-1})(s+\tau_\mathrm{r}^{-1})-\frac{r_\mathrm{AGB}}{\tau_\star\tau_\mathrm{r}}\Big] = \frac{\mathrm{e}^{-t_\mathrm{D}s}m_{x,0}\Gamma(s)}{\tau_\mathrm{p}}\frac{s+\tau_\mathrm{r}^{-1}}{s+\tau_\mathrm{p}^{-1}},
\label{eqn::recycling2}
\end{equation}
where 
\begin{equation}
    \tau_\mathrm{d}=\tau_\star/(1+\eta-r_\mathrm{II})=(\tau_\mathrm{dep}^{-1}-r_\mathrm{II}\tau_\star^{-1})^{-1}.
\end{equation}
We then complete the square to find
\begin{equation}
U_x(s) = \frac{m_{x,0}\Gamma(s)\mathrm{e}^{-t_\mathrm{D}s}(s+\tau_\mathrm{r}^{-1})}{\tau_\mathrm{p}(s+\tau_+^{-1})(s+\tau_-^{-1})(s+\tau_\mathrm{p}^{-1})},
\label{eqn::tau_plusminusp}
\end{equation}
where
\begin{equation}
\tau_{+/-}^{-1} =   \frac{1}{2}(\tau_\mathrm{d}^{-1}+\tau_\mathrm{r}^{-1})\pm\frac{1}{2}\sqrt{(\tau_\mathrm{d}^{-1}+\tau_\mathrm{r}^{-1})^2-4\tau_\mathrm{r}^{-1}(\tau_\mathrm{d}^{-1}-r_\mathrm{AGB}\tau_{\star}^{-1})}.
\end{equation}
This expression for $U_x$ can be separated into individual factors of $(s+\tau^{-1})^{-1}$ using the identity
\begin{equation}
(s+\tau_i^{-1})\prod_j\frac{1}{s+\tau_j^{-1}}=\sum_j\prod_{k\neq j}\frac{\tau_{k,j}}{\tau_{i,j}}\frac{1}{(s+\tau_j^{-1})}.
\label{eqn::factors}
\end{equation}
Equation~\eqref{eqn::tau_plusminusp} then becomes
\begin{equation}
U_x(s) = \frac{m_{x,0}\Gamma(s)\mathrm{e}^{-t_\mathrm{D}s}\tau_{-,+}}{\tau_\mathrm{p}(s+\tau_\mathrm{p}^{-1})}\Big[\frac{\tau_{r,+}^{-1}}{s+\tau_+}-\frac{\tau_{r,-}^{-1}}{s+\tau_-}\Big],
\label{eqn::recycling_soln}
\end{equation}
giving the solution for the mass fraction as
\begin{equation}
Z_x(t) = \frac{\tau_{-,+}}{\tau_{r,+}}Z_x^0(t;\tau_\mathrm{d}=\tau_+)+\frac{\tau_{+,-}}{\tau_{r,-}}Z_x^0(t;\tau_\mathrm{d}=\tau_-),
\end{equation}
where $Z_x^0(t;\tau_\mathrm{d}=\tau)$ is the metallicity-independent yield solution for a depletion time of $\tau_\mathrm{d}=\tau$ (e.g. equations~\eqref{eqn::general_z} and~\eqref{eqn:leq_general_z}). Increasing $\tau_r$ acts to add an additional delay time of return (approximately $\tau_+$) in a similar way to the inclusion of the warm phase in Appendix~\ref{appendix::warm} leading to slower build-up of elements whilst increasing $r_\mathrm{AGB}$ increases the depletion time (approximately $\tau_-$) as more mass is returned leading to a slower depletion rate. Note that due to the linearity, we can still sum solutions for different channels in the same way as in the main body. 

If we consider the case $\tau_\mathrm{r}\rightarrow0$, i.e. all recycling is instantaneous, we see from equation~\eqref{eqn::tau_plusminusp} that $\tau_+\rightarrow\tau_r$ and $\tau_-\rightarrow(\tau_\mathrm{d}^{-1}-r_\mathrm{AGB}\tau_\star^{-1})^{-1}$ such that equation~\eqref{eqn::tau_plusminusp} becomes
\begin{equation}
U_x(s) = \frac{m_{x,0}\Gamma(s)\mathrm{e}^{t_\mathrm{D}s}}{\tau_\mathrm{p}(s+\tau_\mathrm{p})(s+\tau_\mathrm{d}^{-1}-r_\mathrm{AGB}/\tau_\star)},
\end{equation}
i.e. a standard delayed production but now with the depletion time given by $\tau_\star/(1+\eta-r)$. This is exactly the `instantaneous' recycling solution presented by \cite{Weinberg}. In Appendix~\ref{appendix::warm} we briefly consider the case when $x=y$ in the presence of a warm ISM reservoir that adds an additional cooling timescale to the equations. This results in solving a cubic making the solutions for the equations less attractive.

In Fig.~\ref{fig:apdx}, we compare the default model with instantaneous recycling ($\tau_\mathrm{d}=\tau_\star/(1+\eta-r)$) shown in Fig.~\ref{fig:example_models} with the model from equation~\eqref{eqn::recycling_soln} with $r_\mathrm{AGB}=0.3$ and $\tau_\mathrm{r}=3\,\mathrm{Gyr}$. Instantaneous recycling leads to increased yields relative to the delayed recycling (partly artificially driven by the exponential delay-time distribution underprediction of early recycling). At later times, the delayed recycling model `catches up'. The difference in the models is small (as found in figure 7 of \cite{Weinberg})

In conclusion, we find that additional dependence of the return of an element $x$ on its own mass fraction leads to an increase in the depletion time of element $x$ resulting in a slower build-up and slower approach to the equilibrium abundance, which can be approximated as modifying the mass loading factor as $\eta\leftarrow\eta-r$. This effective $\eta$ is used in all models in the main body.

\begin{figure}
    \centering
    \includegraphics[width=\columnwidth]{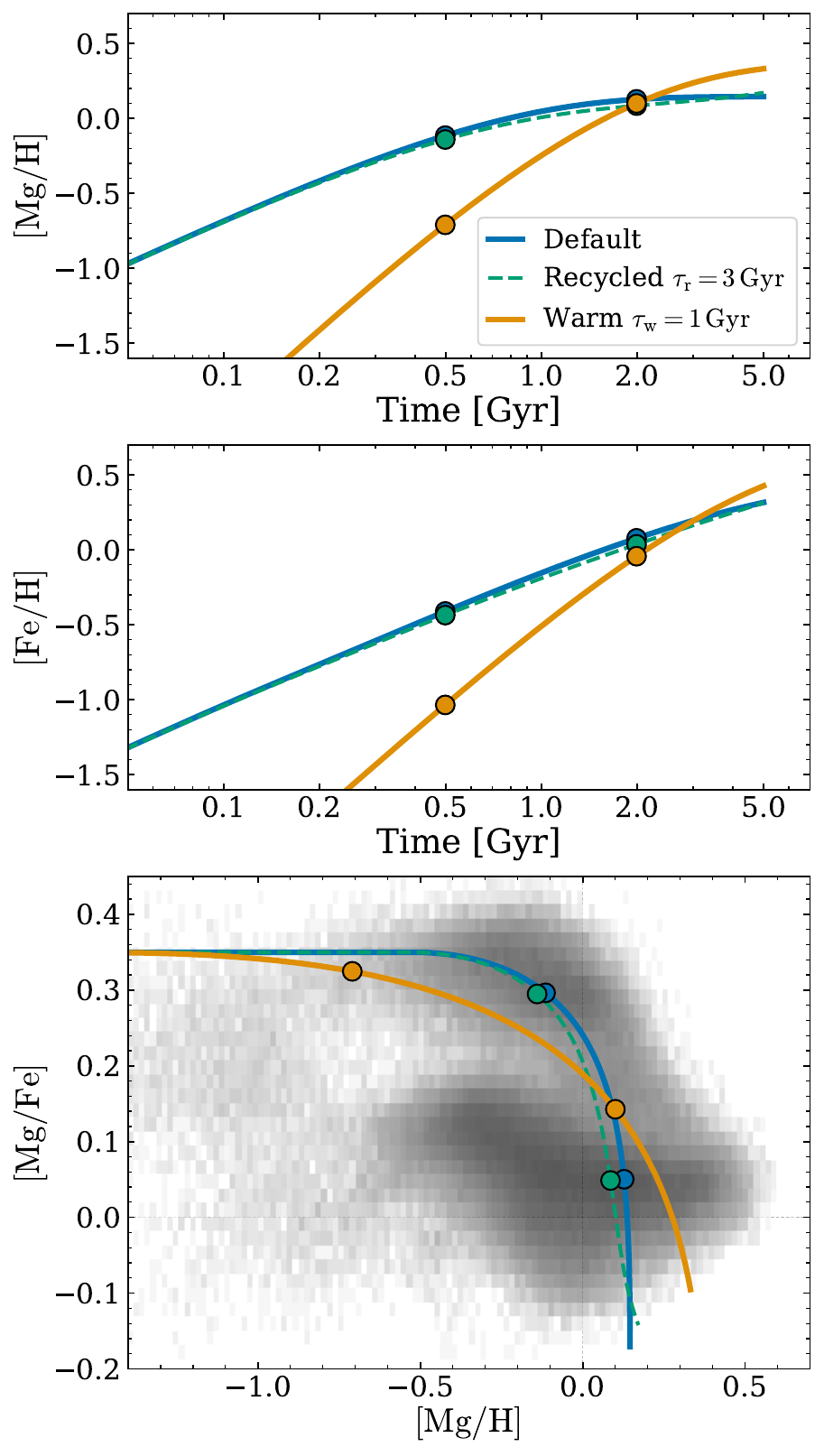}
    \caption{Example models: top two panels show [Mg/H] and [Fe/H] vs. time and the bottom panel shows [Mg/Fe] vs. [Mg/H] with the density of the APOGEE data shown in greyscale. The three models are: 1. `Default' (blue) which is the default exp. model from Fig.~\ref{fig:example_models}, 2. `Recycled' (green dashed) which recycles elements from AGB stars on a timescale $\tau_\mathrm{r}=3\,\mathrm{Gyr}$ and 3. `Warm' (orange) which passes all produced elements through a warm ISM phase which cools back into the star-forming ISM on a timescale of $\tau_\mathrm{w}=1\,\mathrm{Gyr}$.}
    \label{fig:apdx}
\end{figure}

\section{Incorporating a warm ISM phase}
\label{appendix::warm}
In Section~\ref{section:model}, we consider the case when elements are delayed in their return to the ISM by a single process that could be modelled with an exponential delay-time distribution. This could represent either delay due to stellar evolution e.g. in the case of Type Ia supernovae or neutron star mergers, or delay via a warm ISM reservoir that cools into the cold star-forming medium. Here we consider the generalization where the elements are delayed by \emph{both} a stellar evolutionary delay and a warm reservoir \\citep[see][for example chemical evolution models including warm ISM phases]{SchoenrichBinney2009,Chen2023}. In the case the return of the products is delayed by a warm ISM phase with cooling time $\tau_\mathrm{w}$ \citep{SchoenrichWeinberg2019}, equation~\eqref{eqn::gx} becomes
\begin{equation}
\mathcal{G}_x(t)=\int_0^{t}\mathrm{d}t'\frac{1}{\tau_\mathrm{w}}\mathrm{e}^{-(t-t')/\tau_\mathrm{w}}\int_0^{t'}\mathrm{d}t''\,\mathcal{D}(t'-t'')S_x(t'').
\label{eq::gx_warm}
\end{equation}
This is an additional convolution so in the Laplace transform space becomes an additional product such that equation~\eqref{eqn::laplace} becomes
\begin{equation}
U_x(s) = \frac{\mathrm{e}^{-t_\mathrm{D}s}}{\tau_\mathrm{p}\tau_\mathrm{w}}\frac{1}{s+\tau_\mathrm{d}^{-1}}\frac{1}{s+\tau_\mathrm{w}^{-1}}\frac{1}{s+\tau_\mathrm{p}^{-1}}\mathcal{L}_s\Big\{S_x(t)\Big\}.
\label{eqn::u_with_warm}
\end{equation}
Again, let us consider the mass returned as $S_x(t)=m_{x,0}(1+g_xZ_y(t))\sfr(t)$. First, we consider the case when $x=y$ such that the returned mass depends on the initial mass. As discussed in Appendix~\ref{appendix::x_eq_y}, this case is most appropriate for elements that are not significantly produced via a channel and only the initial product is returned. By analogy with equation~\eqref{eqn::recycling1}, we observe that the solution for $U_x(s)$ requires solving a cubic equation for $s$. This is tractable but we do not quote the result here. In the case of prompt return $\tau_\mathrm{p}\rightarrow0$, we return to the results given in Appendix~\ref{appendix::x_eq_y} with the replacement $\tau_\mathrm{p}\leftarrow\tau_\mathrm{w}$.

In the case where the mass of $x$ returned depends on the mass fraction of an element $y$ which arises entirely from a prompt process but which passes through the same warm reservoir (i.e. $\tau_\mathrm{w}$ is the same for both processes), equation~\eqref{eqn::u_with_warm} becomes
\begin{equation}
U_x(s) = \frac{\Gamma(s)\mathrm{e}^{-t_\mathrm{D}s}\tau_\mathrm{w}^{-1}\tau_\mathrm{p}^{-1}}{(s+\tau_\mathrm{d}^{-1})(s+\tau_\mathrm{w}^{-1})(s+\tau_\mathrm{p}^{-1})}\Big(1+\frac{\tau_x^{-1}\tau_\mathrm{w}^{-1}}{(s+\tau_\mathrm{d}^{-1})(s+\tau_\mathrm{w}^{-1})}\Big).
\end{equation}
Again, we observe that  the additional term acts like a derivative with respect to the timescales, such that the additional metallicity-dependent part of the solution $Z^\mathrm{warm}_x(t)$ is
\begin{equation}
Z^{1,\mathrm{warm}}_x(t)=\frac{1}{\tau_x\tau_\mathrm{w}^2}\frac{\partial}{\partial\tau_\mathrm{w}^{-1}}\tau_\mathrm{w}\frac{\partial}{\partial\tau_\mathrm{d}^{-1}}Z^{0,\mathrm{warm}}_x(t),
\label{eqn::warm_operator}
\end{equation}
where $Z^{0,\mathrm{warm}}_x(t)$ is the solution in the metallicity-independent case including a warm reservoir delay. As per the discussion under equation~\eqref{eqn::simple_metal_sum}, the independent parameters are $\tau_\star$ and $\tau_\mathrm{d}$ such that the derivatives are taken at constant $\tau_\star$.

The additional delay from a warm phase can be separated from the stellar evolutionary delay by recognising using the identity in equation~\eqref{eqn::identity1} which gives
\begin{equation}
\frac{1}{\tau_\mathrm{w}\tau_\mathrm{p}(s+\tau_\mathrm{w}^{-1})(s+\tau_\mathrm{p}^{-1})} = \frac{\tau_{\mathrm{p},\mathrm{w}}}{\tau_\mathrm{w}\tau_\mathrm{p}}\Big(\frac{1}{s+\tau_\mathrm{w}^{-1}}-\frac{1}{s+\tau_\mathrm{p}^{-1}}\Big).
\end{equation}
If we explicitly denote the parametrization of $Z_x^0(t)$ for a single exponential delay time with timescale $\tau$ as $Z_x^0(t;\tau)$ (and similarly for $Z^1_x$), we see that the convolution of two timescales can be considered as the sum of solutions each depending on one of the timescales:
\begin{equation}
Z_x^{0,\mathrm{warm}}(t) = \frac{\tau_{\mathrm{p},\mathrm{w}}}{\tau_\mathrm{p}}Z_x^0(t;\tau_\mathrm{w})-\frac{\tau_{\mathrm{p},\mathrm{w}}}{\tau_\mathrm{w}}Z_x^0(t;\tau_\mathrm{p}).
\label{eqn:warm_linearsup}
\end{equation}
The application of equation~\eqref{eqn::warm_operator} then yields
\begin{equation}
Z^{1,\mathrm{warm}}_x(t)=
\frac{\tau_{\mathrm{p},\mathrm{w}}^2}{\tau_\mathrm{w}^2}Z^1_x(t;\tau_\mathrm{p})-\frac{\tau_{\mathrm{p},\mathrm{w}}}{\tau_\mathrm{p}\tau_\mathrm{w}^2}\frac{\mathrm{d}}{\mathrm{d}\tau_\mathrm{w}^{-1}}\tau_\mathrm{w}Z^1_x(t;\tau_\mathrm{w}).
\label{eqn::warm_linearsup1}
\end{equation}
Note that these parametrized solutions, e.g. $Z_x^0(t;\tau)$, are distinct from the parametrized solutions in Appendix~\ref{appendix::x_eq_y},  $Z_x^0(t;\tau_\mathrm{d}=\tau)$. The former changes the delay time, $\tau_\mathrm{p}$, whilst the latter changes the depletion time, $\tau_\mathrm{d}$. In Fig.~\ref{fig:apdx}, an example model with a warm phase is included: the default exponential star formation rate model from Fig.~\ref{fig:example_models} with $\tau_\mathrm{w}=1\,\mathrm{Gyr}$. The inclusion of a warm phase leads to further delay in the return of products to the star-forming ISM and hence the build-up of metals in these solutions is slower. At early times, the additional delay leads to a further factor of time in the solutions so for the exponential star forming case, the early time behaviour is $Z_x^0(t)\propto (\Delta t)^3$. The `knee' in [Mg/H] vs. [Mg/Fe] is more gradual. At late times, the model with the warm phase ends up overtaking the default model as the products are injected at later times into a lower mass gas reservoir leading to enhanced mass fractions.
For added realism and complexity, we might consider a fraction $f_\mathrm{w}$ of products that first enter the warm phase and the remainder $1-f_\mathrm{w}$ enter straight into the cold phase.

\section{Alternative functional forms for the delay-time distribution}
\label{appendix::alternative_functional_forms_for_the_delay_time_distribution}
In Section~\ref{section:model}, we assumed an exponential delay-time distribution. As recently highlighted by \cite{Palicio2023}, two other astrophysically interesting and mathematically tractable delay-time distributions have analytic solutions in our framework. We quote the simplest result assuming metallicity-independent yields and an exponential star formation history $\dot M_\star(t)\propto\mathrm{e}^{-t/\tau_\mathrm{s}}$, but note that the more general models can be derived from these through differentiation with respect to model parameters (see Section~\ref{sec::generalizations}). 

First, we consider the Gaussian delay-time distribution
\begin{equation}
    \mathcal{D}(t) = A \mathrm{e}^{-(t-t_0-t_\mathrm{D})^2/2\sigma^2}\Theta(t-t_\mathrm{D}),
\end{equation}
where we have introduced the normalisation constant $A$ to simplify the presentation. The Laplace transform is
\begin{equation}
\mathcal{L}_s\{\mathcal{D}(t)\}=\sqrt{\frac{\pi}{2\sigma^2}}\mathrm{exp}(\tfrac{1}{2}s^2\sigma^2-s(t_0+t_\mathrm{D}))\mathrm{erfc}\Big(\frac{s\sigma^2-t_0}{\sqrt{2}\sigma}\Big),
\end{equation}
resulting in
\begin{equation}
\begin{split}
    M_x(t)=&
\frac{A m_{x,0} \sigma}{\tau_{\mathrm{s},\mathrm{d}}}\sqrt{\frac{\pi}{2}}\mathrm{e}^{-(t-t_\mathrm{D}-t_0)/\tau_\mathrm{d}}\mathrm{e}^{\tfrac{1}{2}\sigma^2/\tau_\mathrm{d}^2}\times\\&\Big[\mathrm{erf}\Big(\frac{\sigma^2/\tau_\mathrm{d}+t_0+t_\mathrm{D}-t}{\sqrt{2}\sigma}\Big)-\mathrm{erf}\Big(\frac{\sigma^2/\tau_\mathrm{d}+t_0}{\sqrt{2}\sigma}\Big)\Big]+(\tau_\mathrm{d}\leftrightarrow\tau_\mathrm{s}),
\end{split}
\end{equation}
for $t>t_\mathrm{D}$ otherwise $0$. Here, erf is the error function.

Secondly, we have the power-law delay-time distribution
\begin{equation}
\mathcal{D}(t) = \frac{B}{t}\Theta(t-t_\mathrm{D}),
\end{equation}
where we have introduced a normalisation constant, $B$. This delay-time distribution has infinite weight so one must choose via $B$ the time interval over which to normalise $\mathcal{D}(t)$. We note that $\mathcal{L}_s\{\mathcal{D}(t)\}=-\mathrm{Ei}(-t_\mathrm{D}s)$ where $\mathrm{Ei}(x)\equiv \int_{-\infty}^x (e^t/t)\mathrm{d}t$ is the exponential integral function implemented in standard numerical mathematical libraries. This results in the solution
\begin{equation}
    M_x(t) = \frac{B m_{x,0}}{\tau_{\mathrm{s},\mathrm{d}}}\mathrm{e}^{-t/\tau_\mathrm{d}}\Big[\mathrm{Ei}\Big(\frac{t}{\tau_\mathrm{d}}\Big)-\mathrm{Ei}\Big(\frac{t_\mathrm{D}}{\tau_\mathrm{d}}\Big)\Big]+(\tau_\mathrm{d}\leftrightarrow\tau_\mathrm{s}),
\end{equation}
for $t>t_\mathrm{D}$ otherwise $0$.

Both of these alternatives involve more complicated special functions than the solutions presented in the main body (error function and exponential integral). However, most numerical libraries will provide fast, robust implementations of these functions giving these alternatives some utility.

\bsp	
\label{lastpage}
\end{document}